%% file: versioned-cfg-editing-paper.tex
\documentclass[sigplan,10pt]{acmart}
\renewcommand\footnotetextcopyrightpermission[1]{}
\include{config}

\mathchardef\mhyphen="2D

\begin{document}

\title{Profile-Guided, Multi-Version Binary Rewriting}

\author{Xiaozhu Meng}
\email{xm13@rice.edu}
\affiliation{
  \institution{Department of Computer Science, Rice University}  
  \city{Houston}
  \state{TX}
  \country{USA}
  \postcode{77005}
}

\author{Buddhika Chamith}
\affiliation{
  \institution{Indiana University}
  \city{Bloomington}
  \state{IN}
  \country{USA}
  \postcode{47405}
}

\author{Ryan R. Newton}
\affiliation{
  \institution{Indiana University}
  \city{Bloomington}
  \state{IN}
  \country{USA}
  \postcode{47405}  
}


\begin{abstract}

The static instrumentation of machine code, also known as binary rewriting, is a power technique,
but suffers from high runtime overhead compared to compiler-level instrumentation.
Recent research has shown that tools can achieve near-to-zero overhead
when rewriting binaries (excluding the overhead from the application specific instrumentation).
However, the users of binary rewriting tools often have difficulties in understanding
why their instrumentation is slow and how to optimize their instrumentation.

We are inspired by a traditional program optimization workflow,
where one can profile the program execution to identify performance hot spots,
modify the source code or apply suitable compiler optimizations,
and even apply profile-guided optimization.
We present profile-guided, \textit{Multi-Version Binary Rewriting} 
to enable this optimization workflow for static binary instrumentation.
Our new techniques include three components.
First, we augment existing binary rewriting to support call path profiling;
one can interactively view instrumentation costs
and understand the calling contexts where the costs incur.
Second, we present \textit{Versioned Structure Binary Editing},
which is a general binary transformation technique.
Third, we use call path profiles to guide the application of binary transformation.

We apply our new techniques to shadow stack and basic block code coverage.
Our instrumentation optimization workflow helps us identify several opportunities
with regard to code transformation and instrumentation data layout.
Our evaluation on SPEC CPU 2017 shows that the geometric overhead of shadow stack and block coverage
is reduced from 7.6\% and 161.3\% to 1.4\% and 4.0\%, respectively.
We also achieve promising results on Apache HTTP Server,
where the shadow stack overhead is reduced from about 20\% to 3.5\%.

\end{abstract}





\maketitle
\fancyhead{}

\section{Introduction}

Binary rewriting is a powerful technique that statically instruments the machine code of an application
to monitor and collect data from its execution.
It can be used to add security checks~\cite{dang2015performance, Zhang2013CFI, vanderVeen2016ATC, vanderVeen2015PCC},
insert performance measurements code to analyze performance~\cite{WilliamsKing2019CodeMason,Welton2019Diogenes,Welton2020IRP},
and assess software correctness~\cite{Gu2018DDR, Khadra2020EBL}.

While powerful, binary rewriting suffers from high runtime overhead compared to compiler-level instrumentation.
We classify the runtime overhead incurred by binary rewriting into two types:
\textit{tool overhead}, which refers to overhead incurred by binary rewriting tools to support instrumentation,
and \textit{application overhead}, which refers to the overhead incurred by application specific instrumentation.
Recent research on binary rewriting focuses on reducing tool overhead
and has achieved near-to-zero tool overhead in many cases~\cite{Dinesh2020RetroWrite, WilliamsKing2020Egalito, Meng2021ICFGP}.

On the other hand, application overhead is optimized by binary rewriting application researchers,
who design drastically different optimization strategies tailored to their own applications~\cite{Gu2018DDR, Khadra2020EBL}.
These optimization strategies are often the results of numerous, trial-and-error attempts.
We believe it is valuable to have a systematic approach for optimizing binary rewriting applications.

A traditional optimization approach for software programs is to
profile the program to identify performance hot spots and
transform the hot spots either through manual source code modification or tailored compiler optimizations.
In addition, profile-guided optimizations have been shown to be effective for general program optimization~\cite{Panchenko2019BOLT}
and even for compiler-level instrumentation~\cite{Duta2021PIBE}.

Unfortunately, users of binary rewriting tools do not have aforementioned powerful weapons at their disposal.
First, rewritten binaries generated by existing binary rewriting tools are not suitable for call path profiling~\cite{Hall1992CPP}.
Profiling tools such as HPCToolkit~\cite{HPCToolkit2010Adhianto} use call path profiling
to attribute performance metrics, including running time, cache misses, and page faults, to program calling contexts.
Users can know the specific functions, loops or statements that incur the most costs.
Without call path profiling, users may know the overall inefficiency in their instrumentation,
but cannot pinpoint the specific program contexts to design optimization strategies.

Second, existing binary rewriting tools are unable to perform
necessary binary transformation to support instrumentation optimization.
Compilers can drastically change program structures to perform a variety of optimizations,
including function inlining and loop unrolling.
While we should not expect a binary rewriting tool to have the strength of a compiler
with regard to program transformations,
we believe there is a middle ground where useful optimizations can be achieved at the binary level.

In this paper, we present \textit{multi-version binary rewriting (MVBR)} to fill this void.
The basic idea of MVBR is to create multiple copies of a piece of code and
support arbitrary control flow redirection among copied code, original code, and instrumentation.
We can instrument these different copies in different ways (including not instrumenting some copies).
``a piece of code'' can refer to any programming language code constructs available at the binary level,
including functions, loops, and basic blocks.
This capability enables us to drastically transform the binary and customize instrumentation according to its invoking contexts.

We design \textit{Versioned Structure Binary Editing (VSBE)} to support the binary transformations used by MVBR.
VSBE provides a set of primitive binary transformation operations,
including cloning a basic block, redirecting a control flow edge, and splitting a basic block.
The composition of these operations can lead to power binary transformation.

MVBR also generates rewritten binaries that are suitable for call path profiling.
Application researchers can interactively view performance metrics attributed to source code,
investigate the conditions where the instrumentation incurs high overhead,
and design new instrumentation strategies accordingly.
In addition, we leverage profiles to guide where to apply MVBR,
as making copies of code blindly will cause negative effects on the instruction cache.
Profiled-guided MVBR makes it possible to perform familiar, powerful optimizations, 
such as function inlining, at the binary level.
For profiled-guided MVBR, we directly use performance metrics collected at the binary level,
so debug information or symbol tables is not needed.

To demonstrate the effectiveness of our new techniques,
we apply MVBR to two typical binary rewriting applications
and show how instrumentation profiling can reveal optimization opportunities
and how profile-guided MVBR can materialize these opportunities.

The first one is shadow stack~\cite{dang2015performance},
which is a software security policy to mitigate return based attacks.
It works by creating a thread-local shadow stack and inserting instrumentation
that saves return addresses to the shadow stack at a function entry,
and validating the actual return address on the original stack with
the one from the shadow stack before function exits.
The second one is basic block code coverage,
which is a building block for software correctness assessment and fuzzing.
The instrumentation allocates a new region of memory,
where each byte of the memory represents whether an instrumented block has been executed or not,
and inserts code that instrumented blocks to write the execution information.
We choose these two applications because 
(1) they instrument at different code granularity (function vs basic block),
and (2) they are from different application domains.

For both applications, we take the same optimization workflow:
(1) implement a baseline instrumentation,
(2) profile it to observe its performance characteristics and identify optimization opportunities,
and (3) use profile-guided MVBR to realize the optimizations.

For shadow stack, among several optimizations, we highlight \textit{binary function inlining}.
An inlined callee does not need instrumentation as there is no return address any more.
The original callee remains instrumented as we do not necessarily want to inline all calls to function.
Existing work on function inlining~\cite{Scheifler1977AAIS, Chang1989IFE, Ayers1997AI, Arnold2000ACS}
focuses on when to inline a function and operates at the compiler level.
We will present a new algorithm based on VSBE to inline functions at the binary level,
which has to address several additional challenging regarding stack pointer adjustment and tail calls.

For block coverage, we identify performance issues caused by false sharing when running with multiple threads,
hot instrumentation inside loops, and instrumentation data allocation that causes inefficient data cache utilization.
We design a new instrumentation instruction sequence, profile-guided loop transformation,
and profile-guided instrumentation data allocation to address these problems.

We evaluate our work using SPEC CPU 2017.
For shadow stack, 
our optimizations reduced the maximal overhead from 31.0\% to 10.0\%
and the (geometric) mean overhead from 7.6\% to 1.4\%.
For block coverage,
our optimizations reduced the overhead from over 100\% to
12.1\% maximal and 4.0\% mean overhead.
We also our optimizations to two real world software,
including Apache HTTP Server,
where the shadow stack overhead is reduced from over 20\% to about 3\%.

In summary, this work makes the following contributions:

\begin{itemize}
\item \textit{Multi-Version Binary Rewriting}, a new binary-rewriting instrumentation technique 
that supports customizing instrumentation by transforming binaries,
\item \textit{Versioned Structure Binary Editing}, a new binary transformation technique 
as the foundation of multi-version binary rewriting,
\item support for call path profiling in rewritten binaries, which enables interactive instrumentation performance 
analysis and profile-guided multi-version binary rewriting,
\item instrumentation profiling analysis and new optimizations to improve shadow stack and basic block code coverage.
\end{itemize}

\section{Related Work}

We put our work in context by surveying related work of binary rewriting tools, profile-guided optimizations,
and application specific research that uses binary rewriting.

\subsection{Binary Rewriting Tools}

Binary rewriting tools support inserting arbitrary code at arbitrary locations in the original binary.
There are two main approaches to achieve this goal.

The first approach is code patching, 
which patches original code with trampolines (typically branch instructions)
to redirect control flow from original code to a new area of code where instrumentation is inserted.
Tools using this approach include Dyninst~\cite{Bernat2011AWAT, Bernat2011ESR, Meng2021ICFGP} and E9Patch~\cite{Duck2020BRW}.

E9Patch~\cite{Duck2020BRW} uses instruction patching, which modifies individual instrumented instructions to branch to the new area of code.
The new area of code contains only the instrumentation, the modified instruction, and a branch back to the original code.
Instruction patching does not require any binary analysis and thus enjoys high generality.
However, it incurs high tool overhead as the control flow may frequently bounce between original and instrumented code.

Incremental CFG Patching~\cite{Meng2021ICFGP}, implemented as an extension to the mainstream Dyninst,
significantly reduces tool overhead by relocating all instrumented functions to the new area of code,
modifying direct control flow to stay in relocated code area,
and opportunistically modifying indirect control flow including jump tables and function pointers.
Incremental CFG patching can achieve near-to-zero tool overhead
and provides a failure mode analysis when the underlying binary analysis failed.

The second approach is IR lowering, which first lifts the binary to a low-level IR and then re-generate the rewritten binary.
Tools in this category lift binaries to either tool specific IR~\cite{WilliamsKing2020Egalito, Hawkins2017Zipr}
or the assembly language~\cite{wang2017ramblr,Montoya2020DD, Dinesh2020RetroWrite}.
Recent IR lowering tools enjoy near-to-zero tool overhead.
However, this approach relies on complete binary analysis to be able to fully lift a binary.
Tools either rely on runtime relocation entries~\cite{WilliamsKing2020Egalito, Dinesh2020RetroWrite}
or binary analysis heuristics without clear failure mode analysis~\cite{Hawkins2017Zipr, Montoya2020DD}.

We observe that neither approaches can leverage optimization passes implemented in modern compilers.
Such optimization passes in compilers typically work at a higher-level IR (such as LLVM IR).
Our work is orthogonal to the research of binary rewriting tools.

\subsection{Profile-Guided Optimization}

Profile-Guided Optimization (PGO)
has been successful in improving program performance~\cite{Cohn1998OAE, Ottoni2017OFP, Chen2016AutoFDO, Chen2013THE}.
Representative PGO tools include Facebook BOLT~\cite{Panchenko2019BOLT, Panchenko2021LBOLT} and Google Propeller~\cite{PROPELLER}.

Our work leverages profiles to optimize instrumentation and differs from existing PGO research on the following two aspects.
First, we work fully at the binary level, requiring no recompilation or relinking.
While BOLT also performs optimization at binary level, BOLT requires link-time relocations to be present in the binary.
Users of BOLT often have to recompile the program with flag \code{-Wl,-q} to 
instruct the linker to retain link time relocation entries.
Google Propeller also requires recompilation to perform PGO.

Second, existing PGO focuses on improving code and data layout to improve instruction and data cache efficiency.
In contrast, our work uses profiles to guide binary transformation to elide instrumentation.
In other words, our work uses PGO to improve instrumentation policies, while existing PGO research focuses
on improve general program execution efficiency.
Orthogonally, we can leverage existing code layout PGO policies to further improve overall performance.

Duta et al.~\cite{Duta2021PIBE} uses profiles to elide instrumentation to reduce the overhead of 
the defenses for transient execution attacks at the compiler level.
Compared to this work, we work at the binary level and provides a general instrumentation optimization strategy

\subsection{Binary Rewriting Applications}

We discuss the 
state-of-the-art implementations of shadow stack and block coverage at the binary level,
which will serve as the baselines in this work.

\subsubsection{Shadow Stack}

Shadow stacks prevent control flow hijacking attacks 
which are based on modifying a return address on the call stack.
Shadow stacks maintain a copy of the regular stack in a tamper-proof region, and check that both copies match on all function returns.
This scheme is effective against code reuse attacks like Return-oriented programming (ROP) \cite{roemer2012return} and
return-to-libc~\cite{wojtczuk2001advanced}.

The basic implementation of shadow stack instruments every function entry and exits.
Naturally, one can skip instrumenting a function if the function will never overwrite any return address.
The idea of using program analysis to reduce instrumentation is often leveraged by compiler level instrumentation.
For example, gcc's \code{-fstack-protector} flag will insert stack guard for functions that call \code{alloca}, 
and functions with buffers larger than 8 bytes.

We find that there is no existing work that utilize 
static binary analysis to improve shadow stack
done at the binary level.
Therefore, for shadow stack, the baseline used in this work instruments every function.

\subsubsection{Block Coverage}

Code coverage determines how much and which piece of code is executed for a given input.
We use block coverage as the example. Our optimization strategies can be extended to edge coverage.

A naive implementation is to instrument every basic block.
Previous research showed that we can reduce instrumentation
by performing dominator analysis on
the control flow graph of a function~\cite{Agrawal1994Coverage}.
This optimization has been shown to be also effective at binary level~\cite{Khadra2020EBL},
which will be the baseline for block coverage in our work.

\section{Versioned Structure Binary Editing}

VSBE improves the seminal work of Structure Binary Editing (SBE), presented by Bernat and Miller~\cite{Bernat2012SBE}.
SBE defines a set of binary transformation operations, including block cloning, control flow edge redirection, and function cloning.
However, SBE does not support redirecting indirect control flow that uses jump tables,
which are often used by compilers to generate switch statements in C/C++.
In addition, transformation operations provided by SBE do not always compose,
which limits its capability to design complex binary transformations.
In this section, we describe how VSBE to address these weaknesses.

\subsection{Definitions}

We build upon the definitions used by SBE.

\textbf{Control Flow Graph:} We start with a standard CFG definition, $G=<B, E, F>$, where $B$ is a set of basic blocks,
$E$ is a set of control flow edges between the basic blocks, and $F \subseteq B$ is a set of entry blocks of functions.

\textbf{Basic Block:} A basic block, $b=<s, t, v, jt>$, represents a sequence of machine instrumentations that have incoming
control flow only at its entry and outgoing control flow only at its exit.
$s$ and $t$ represent the start and end address of a basic block.

We add a version number $v$ to the definition of a basic block.
Given an original block $b$ starting at address $s$,
all clones of this block will have the same starting address $s$.
Therefore, it is necessary to have a distinct version number for each of them.
Conventionally, the version number for all original blocks are $0$.
The version numbers of cloned blocks should be set based on certain semantic grouping.
For example, when we clone a loop, all basic blocks inside the loop should have the same version number.
In \cref{sec:multi-version}, we will present examples of setting version numbers.

We also add $jt$ to represent a potential jump table used by a basic block.
If a basic block does not use a jump table to compute indirect jump targets, $jt$ is empty.
Otherwise, $jt=<ts, te, stride, slice>$, where $ts$ is the start address of the jump table;
$te$ is the end address; $stride$ is the size of a jump table entry; $slice$ is a backward slice from the indirect jump
that includes all instructions needed for computing the control flow target.
The values for each element in $jt$ can be calculated during CFG construction,
which is a capability provided by many modern binary analysis tools~\cite{Angr2016, Meng2016BinaryNotEasy, DiFederico2017RUB}.

\textbf{Control Flow Edge:} A control flow edge, $e=<b_s, b_t, type>$,
represents control flow transfer from a source block $b_s$
to a target block $b_t$, annotated with a transfer type $type$.
In SBE, edge types include direct, conditional-taken, condition-not-taken, fall-through.
We reduce conditional-not-taken, fall-through all to direct.
The reason for this change is that after making multiple copies of a code,
copies of a fall-through edge will not be fall-through any more.

\textbf{Function:} A function is defined as a set of basic blocks that is reached by 
traversing only intra-procedural edges from the entry block.
This definition for a function can cope with challenging cases where
functions share basic blocks and functions have non-contiguous basic blocks.

\subsection{Primitive CFG Transformation Operations}

SBE defined several primitive CFG transformation operations
as the foundation for performing binary transformation.
We discuss how we improve these operations in VSBE.

\begin{figure*}  
  \centering
  \begin{subfigure}[b]{0.15\textwidth}
    \centering
    \includegraphics[page=1, width=\textwidth]{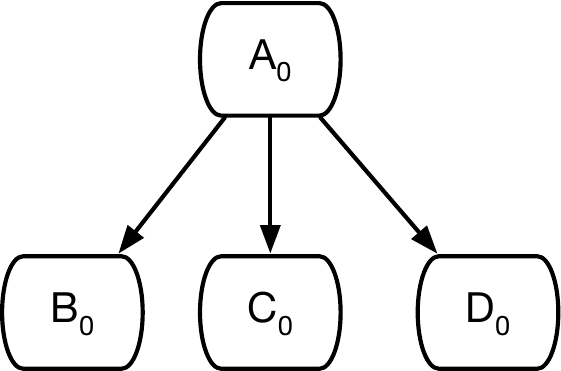}  
    \caption{Original CFG}
    \label{fig:primitive:op:orig}
  \end{subfigure}
  \hfill
  \begin{subfigure}[b]{0.35\textwidth}
    \centering
    \includegraphics[page=2, width=\textwidth]{figures/cfg-primitive-op.pdf}  
    \caption{Clone blocks}
    \label{fig:primitive:op:bc}
  \end{subfigure}
  \hfill
  \begin{subfigure}[b]{0.35\textwidth}
    \centering
    \includegraphics[page=3, width=\textwidth]{figures/cfg-primitive-op.pdf}  
    \caption{Redirect indirect jump edges}
    \label{fig:primitive:op:er}
  \end{subfigure}
  \caption{An example of cloning a function.}  
  \label{fig:primitive:op}
\end{figure*}

\cref{fig:primitive:op} shows an example of cloning a functions.
The steps of cloning a loop are similar.
\cref{fig:primitive:op:orig} shows the original CFG of the example function.
This function has four basic blocks. The entry block A has an indirect jump,
whose targets include block B, C, and D. Block B, C, and D are return blocks. The subscripts after the block name represents its version
number. Blocks with version number 0 are original code. Blocks with version number 1 are copied code.

\textbf{Block Clone}: In SBE, cloning a basic block $b$ includes cloning the block itself and also cloning the outgoing edges of the block.
The cloned block does not have any incoming edges.
Users of SBE must perform edge redirection to create incoming control flow to the new block.
As shown in \cref{fig:primitive:op:bc}, we clone all blocks in the function.
Outgoing edges of block $A_1$ remain targeting $B_0$, $C_0$, and $D_0$.

\textbf{Edge Redirection}:
Conceptually, edge redirection is simple to do as we only need to update the target block of an edge.
However, SBE does not support redirecting indirect edges.
The targets of indirect edges are determined through runtime computation.
To redirect indirect edges, we must modify the computation of the indirect edge targets.
VSBE support redirecting indirect edges by cloning jump tables and modifying jump target calculation.
As shown in \cref{fig:primitive:op:er}, we redirect the targets of the new edges from
$B_0$, $C_0$, and $D_0$, to $B_1$, $C_1$, and $D_1$, respectively.

\textbf{Block Split}:
SBE defines that we can split a block by choosing an address inside the block range.
The result of block split is that the original block is shrunk,
a new block is created at the split address,
and a new edge is created from the shrunk block pointing to the new block.

The newly created edge can then be redirected to other blocks, which makes
the second half of the block dead code.
This achieves the goal of removing instructions.

\subsection{Code and Data Generation}

For a basic block $b=<s, t, v, jt>$, we copy the instructions in the address range $[s, t)$ from the original binary
and update every instruction that uses PC-relative addressing to the same global data is referenced.
During this step, if $jt$ is not empty, we do not emit new a jump table as the addresses of the indirect jump targets
in the rewritten binary may have not been set.

When generating a control flow edge,
for direct edges and condition-taken edges,
we generate a direct branch and a conditional branch, respectively.
In addition, we need to compute the PC-relative displacement based on the
start address of the target block.

Finally, we emit new jump tables and update the instructions that reference them.
Suppose we need to update a block $b_1=<s_1, t_1, v_1, jt>$ and $jt=<ts, te, stride, slice>$ to use a new jump table.
The new jump table $jt'=<ts', te', stride, slice>$ is allocated with the following property.
First, $ts'$ will be the first available virtual address to hold the new table and 
$ts'$ is aligned based on the table stride to avoid access violation.
$te'$ can be computed by adding the table size to $ts'$.
The contents of the jump table are determined by target blocks,
which may have been redirected and are different from the original targets.
If the entry of a target block is instrumented, the target address should be the beginning of instrumentation.
Otherwise, instrumentation may be skipped.
We also need to update the instruction that computes the location of the jump table to reference the new table.
From $slice$, we can determine the original instruction and its belonging basic block that performs the computation.
Finally, we use the version number $v_1$ to lookup the correct clone that contains the instruction to update.

\section{Multi-Version Instrumentation}
\label{sec:multi-version}

We show how to compose primitive binary transformation operations
provided by VSBE to optimize static binary instrumentation.

\subsection{Binary Function Inlining}

Function inlining is a common compiler optimization that removes the costs of function calls and returns.
In the context of binary instrumentation,
binary function inlining can bring additional benefits.
First, for shadow stack, an inlined call no longer needs instrumentation
as there is no return address.
Second, function inlining naturally extends intra-procedural static binary analysis
to be context-sensitive and inter-procedural for the inlined call sites,
and thus increases the utility of static binary analysis for eliding instrumentation.
If we create a copy of the callee, inline it into the caller,
instrument the original copy of the callee in the original way,
we guarantee that the optimization is done only for the inlined call site,
and it will not impact other call sites.

Next, we describe the algorithm for inlining binary functions.
As shown in ~\cref{fig:binary_function_inlining}, it takes a call block $cb$ as input.
Here, we assume that $cb$ makes a direct function call.
For an indirect call, we first do indirect call promotion~\cite{Duta2021PIBE}
that transforms an indirect call site to directly call a subset of the call targets,
and then apply direct call inlining to the promoted calls.

Line 2 - 4 identify and clone the callee for inlining.
At line 5, we utilize procedure \code{SplitAndRedirect} to split the input call block $cb$
before its last instruction (i.e. the call instruction), and redirect the outgoing edge to the cloned function entry block.
At line 6, procedure \code{RedirectReturns} will iterate every cloned return block and use \code{SplitAndRedirect}
to remove its last instruction (i.e. the return instruction), and redirect outgoing control Flow
to the call fall-through block in the caller.

\code{SplitAndRedirect} is shown at line 9 - 18.
If the block to split only has one instruction,
we do not need to split block.
Instead, we need to redirect every incoming control flow edge to the new target,
as shown at Line 11 - 13.
Otherwise, we split the block at its last instruction and redirect outgoing edge to the new target.

\begin{algorithm}  
  \begin{algorithmic}[1]  
  \Procedure{BinaryFunctionInlining}{ $cb$ }
  \State $callee \gets$ \textit{getCallee(cb)} 
  \State $cftb \gets$ \textit{getCallFTBlock(db)}
  \State $cloneEntry, cloneRetBlocks \gets$ \textit{CloneFunction(callee)}
  \State \textit{SplitAndRedirect(cb, cloneEntry)}
  \State \textit{RedirectReturns(cloneRetBlocks, cftb)}
  \EndProcedure  
  \Procedure{SplitAndRedirect}{$b$, $newTarget$}
  \If {\textit{numberOfInstruction(b)} == 1}
    \For {$e \in pred(b)$}
      \State \textit{RedirectEdge(e, newTarget)}
    \EndFor
  \Else
    \State $e \gets $ \textit{SplitBlock(b, lastInstructionAddress(b))}
    \State \textit{RedirectEdge(e, newTarget)}
  \EndIf
  \EndProcedure

  \end{algorithmic}  
  \caption{An algorithm for binary function inlining.}  
  \label{fig:binary_function_inlining}
\end{algorithm}

\textbf{Setting Version Number:}
For each original function, we prepare a version number counter.
When we make a copy of a basic block during inlining,
we increment and use the counter of the original function where the cloned block belongs to.
This strategy can support the case where an original function is inlined multiple times to another function.

\textbf{Stack Pointer Adjustment:}
The removal of call and return instructions means that 
the inlined callee will be executed with a $+8$ offset for the stack pointer
compared to without inlining. This can cause access violation when the callee
stores or loads a 16-byte (or larger) data element using the stack,
as these memory accesses must be aligned at 16-byte (or larger) boundary.
Note that function inlining at the compiler level does not need to worry this problem
as the compiler will adjust these memory references accordingly.
We address this issue by emitting a ``\code{lea \%rsp, -8}'' instruction before the inlined code
and a ``\code{lea \%rsp, 8}'' instruction after the inlined code to ensure our inlining does not change the stack pointer.

\textbf{Tail Calls:}
Suppose we have function $A$ calls $B$, $B$ tail calls $C$, and we want to inline $B$ into $A$.
Without inlining, $C$ will return to $A$'s call site to $B$.
After performing inlining $B$ into $A$, $C$ cannot return to $A$ anymore as there is no return address.
To handle this issue, we put the address of $A$'s call site to the stack before the tail call in $B$.
In this way, $C$ will correctly return. 
Note that $C$ is still instrumented.

\subsection{Multi-Version Loop Instrumentation}
\label{sec:multi-version:loop}
For idempotent instrumentation, executing it once is equivalent to
executing it multiple times with regard to its semantics.
We observe that such idempotent instrumentation often incurs unnecessary
overhead when it is inside loop.
We present multi-version loop instrumentation to optimize this case.
Without loss of generality, we use block coverage, which is idempotent, as an example in this section.

\begin{figure*}  
  \centering
  \begin{subfigure}[b]{0.15\textwidth}
    \centering
    \includegraphics[page=1, width=\textwidth]{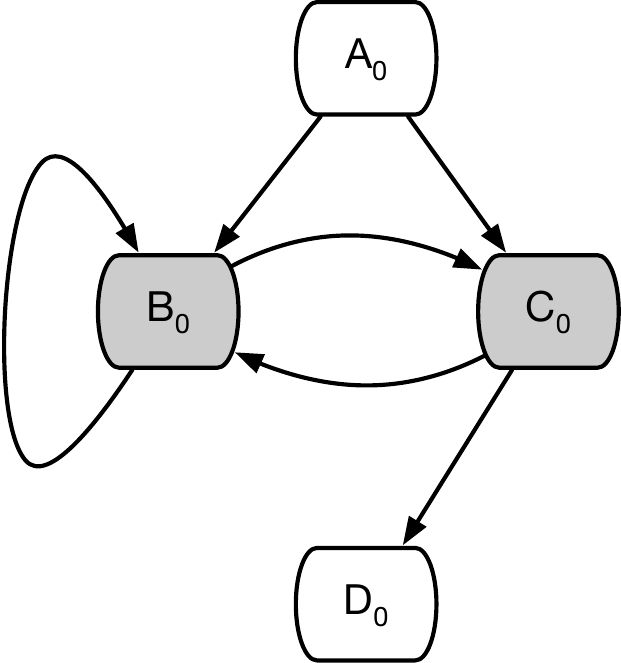}  
    \caption{Original CFG}
    \label{fig:multi-version:loop:orig}
  \end{subfigure}
  \hspace{3em}
  \begin{subfigure}[b]{0.5\textwidth}
    \centering
    \includegraphics[page=2, width=\textwidth]{figures/multi-loop-inst.pdf}  
    \caption{Multi-version loop instrumentation}
    \label{fig:multi-version:loop:clone}
  \end{subfigure}
  \caption{An example of multi-version loop instrumentation to optimize idempotent instrumentation.}  
  \label{fig:multi-version:loop}
\end{figure*}

We start with an example shown in \cref{fig:multi-version:loop} to illustrate the basic idea of multi-version loop instrumentation,
and then describe an algorithm for it.
\cref{fig:multi-version:loop:orig} shows the baseline instrumentation for block coverage.
In this example, we have a function with four basic blocks, denoted as $A_0$, $B_0$, $C_0$, and $D_0$ respectively.
Instrumented blocks are shaded. $B_0$ and $C_0$ are inside a loop and are instrumented.
In this baseline instrumentation, if the loop is executed many times,
instrumentation in $B_0$ and $C_0$ will incur unnecessary overhead, without increasing any coverage information.

\cref{fig:multi-version:loop:clone} shows the results of applying multi-loop instrumentation.
First, we the loop three times,
so we have block $B$ and $C$ with version number from $1$ to $3$.
Different loop versions are instrumented differently:
for version $1$, only $C$ is instrumented, while $B$ is not;
for version $2$, only $B$ is instrumented, while $C$ is not;
for version $3$, neither $B$ nor $C$ is instrumented.

To understand why loop clones are instrumented in the aforementioned way,
we denote a bit vector $<isC, isB>$, where $isB$ ($isC$) is $1$ if and only if the instrumentation for $B$ ($C$)has been executed.
Instrumentation for loop version $1$ represents state $<0, 1>$.
Since instrumentation for $B$ has been executed before, $B_1$ does not need to be instrumented.
And since instrumentation for $C$ has not been executed, $C_1$ is instrumented.
The same reasoning applies to loop version $2$ and $3$.
In addition, if we convert the state bit vector to the corresponding decimal value,
it matches the version number of the loop.

Next, we redirect control flow edges among loop clones. 
The principle is still based on the state bit vector.
For example, for version $0$ whose state vector is $<0,0>$,
if $B_0$ is executed, it means the state has transitioned to $<0,1>$;
therefore, the outgoing edges of $B_0$ should be redirected to the loop representing state $<0, 1>$, which is loop version $1$.
Similar, if $C_0$ is executed, it means the state has transitioned to $<1, 0>$;
therefore, the outgoing edges of $C_0$ should be redirected to loop version $2$.
For targets outside the loop, we do not need to redirect edges.
For example, $C_0$'s outgoing edge to $D_0$ is not redirected.

For blocks inside a cloned loop that are not instrumented,
executing such blocks do not change the instrumentation state.
Therefore, we should not redirect their outgoing edges.
For example, as $B_1$, $C_2$, $B_3$, $C_3$ are not instrumented, their outgoing edges are not changed.

Then, we consider a sample execution trace of the function to compare multi-version loop instrumentation with the baseline.
Suppose we have a block execution trace

\begin{equation}
  \label{trace:baseline}  
  A_0 \rightarrow B_0 ^ N \rightarrow (C_0 \rightarrow B_0) ^ M \rightarrow C_0 \rightarrow D_0 \tag{T1}
\end{equation}

where the superscript means the number of iteration.
In the baseline,
instrumentation for $B$ will be executed $N+M$ times,
and instrumentation for $C$ will be executed $M+1$ times.
In the multi-version loop instrumentation, the execution trace is converted to

\begin{equation}
  \label{trace:new}
  A_0 \rightarrow B_0 \rightarrow B_1 ^ {N-1} \rightarrow C_1 \rightarrow (B_3 \rightarrow C_3) ^ M \rightarrow D_0 \tag{T2}
\end{equation}

Trace \ref{trace:new} and Trace \ref{trace:baseline} are semantically equivalent.
However, as only $B_0$ and $B_2$ contain instrumentation for $B$,
instrumentation for $B$ is executed only once in Trace \ref{trace:new}.
Similarly, as only $C_0$ and $C_1$ contains instrumentation for $C$,
instrumentation for $C$ is also only executed once in Trace \ref{trace:new}.

Now we describe the algorithm for performing multi-version loop instrumentation.
Given a loop $L_0=\{b_1, \ldots, b_n\}$ and a subset of instrumented blocks $inst_0 \subseteq L_0$,
\cref{fig:multi-version-loop:alg} shows the steps for multi-version loop instrumentation.

At Line 2, we calculate the number of copies needed, which is exponential to the size of $inst_0$.
Obviously, this will not scale to large loop where we may need to instrument thousands of blocks.
In \cref{sec:profile}, we will show how to use profiles to identify the top $K$ hot blocks,
and only make loop clones for these hot blocks.

At Line 3 - 5, we clone the loop and set the version number for each clone accordingly.
At this point, the control flow of each clone loop stay in the same version.
The main part of the algorithm is devoted to redirect control flow among copies of loops.
We enumerate every version of loop (Line 6), and
for each version of loop, we enumerate every instrumented block with an index (Line 7)
and denote the corresponding cloned block with $inst_{v, index}$ (Line 8),
where $v$ is the version number and $index$ is the assigned index of the block.

If the $index$th bit of $v$ is set, it means that the corresponding block does not need instrumentation
as the instrumentation has been executed already. 
If not set, we instrument the block (Line 10),
and handle transitions between loop clones by using
procedure \textit{LoopCloneTransition} (Line 11).

\textit{LoopCloneTransition} computes the new version number (Line 18),
and enumerate every outgoing edge of the current block to determine
if we need to redirect this edge (Line 19 - 24).
At Line 20, \textit{BlockLookup(b, v)} returns the block
that has version number $v$ and has the same start address
as input $b$. It is used to lookup a transition target.

\begin{algorithm}
  \begin{algorithmic}[1]  
  \Procedure{MVLoopInstrumentation}{ $L_0$, $inst_0$ }
  \State $copies \gets 2^{|inst_0|}$
  \For {$v \in \{1, \cdots, copies - 1\}$}
    \State $L_v,  inst_v \gets $ \textit{cloneLoop(}$L_0$, $v$\textit{)}
  \EndFor

  \For {$v \in \{0, \cdots, copies - 1\}$}
    \For {$index \in \{0, \cdots, |inst_v|-1\}$}      
      \State $srcb \gets inst_{v, index}$
      \If { \textit{isBitClear(v, index)} }        
        \State \textit{IdempotentInstrumentation(srcb)}
        \State \textit{LoopCloneTransition(} $v$, $index$, $srcb$\textit{)}
      \EndIf
    \EndFor
  \EndFor
  \EndProcedure
  \Procedure{LoopCloneTransition}{$v$, $index$, $srcb$}
  \State $newV \gets $ \textit{setBit(v, index)}        
  \For {$e \in succ(srcb)$}    
    \State $trgb \gets $ \textit{BlockLookup(targetBlock(e), newV)}
    \If {\textit{isInLoop(}$trgb$, $L_{newV}$\textit{)}}     
      \State \textit{RedirectEdge(e, trgb)}
    \EndIf            
  \EndFor  
  \EndProcedure
  \end{algorithmic}
  \caption{An algorithm for multi-version loop instrumentation.}  
  \label{fig:multi-version-loop:alg}
\end{algorithm}

\section{Instrumentation Performance Analysis}
\label{sec:profile}

We augment binary rewriting to support call path profiling
and integrate the support in HPCToolkit~\cite{HPCToolkit2010Adhianto}.

HPCToolkit uses call path profiling.
It interrupts program execution either periodically (using timer interrupts)
or when the event buffer of the Performance Monitoring Unit (PMU) is full.
It then processes corresponding performance events and performs a stack unwind
to associate a sample with the current calling context.

Stack unwinding is often based on frame information encoded in \code{.eh\_frame} sections.
Existing binary rewriting tools do not update \code{.eh\_frame} sections when generating rewritten binaries.
In addition, there is no corresponding frame information for the inserted instrumentation.
These two factors may cause stack unwind to fail for rewritten binaries.

\subsection{Address Mapping Table}

We augment \textit{RA Address Mapping} introduced by Incremental CFG patching~\cite{Meng2021ICFGP} to call path profiling.
RA address mapping instruments the stack unwinding code in language runtime to 
translates the return address on stack back to original return address,
which enables rewriting C++ exceptions and Go binaries.
However, RA address mapping only captures return addresses of call sites in a binary.
During call path profiling, a software interrupt can happen at any address, not just at return addresses.
We augment this mapping to cover all the code.

Each entry in our address mapping table 
represents a contiguous interval in the rewritten binary.
It includes three fields:
the start of the address in the rewritten binary, the length of the interval,
and the start of the address in the corresponding original binary.
For instrumentation code, there is no original corresponding address,
so the third field is set to -1.

We then modify HPCToolkit to load this mapping address table during profiling and 
change its code to translate the PC values.

Our address mapping table is only needed if we want to do instrumentation performance analysis
or perform profile-guided instrumentation. It is not needed in production runs.

\subsection{Alternative Stacks}
\label{sec:profile:as}

Traditionally, instrumentation shares the same stack with the original program.
This leads to several issues that requires additional instructions,
which not only cause runtime overhead but also make it more difficult to perform stack unwinding.
We present a new instrumentation scheme, which uses segment register \code{gs} to hold a pointer
to a memory region allocated at program and thread startup time.

System V ABI specifies a 128-byte red zone space beyond stack pointer.
Program can directly save temporary values in the red zone without allocating a stack frame.
For this reason, to guard against overwriting red-zone,
original stack instrumentation first needs to move down the stack pointer.
Under alternative stack instrumentation,
we do not need to worry about the red-zone space on the original stack.

If the instrumentation includes a function call,
it must conform to the stack pointer alignment requirement specified by the ABI.
While stack pointer can be aligned with a simple \code{and} instruction,
it causes other implications that must be addressed.

First, an \code{and} instruction overwrites flags;
we may have to save and restore flags if the flag register is live at the instrumentation point.
Second, the effect of stack pointer alignment cannot be easily reversed.
This is contrary to moving down stack pointer with a constant offset,
where we can move up the stack pointer with the same constant to reverse its affect.
For this reason, we save \code{\%rsp} to a scratch register,
align the stack pointer, and save the original stack pointer to the current stack top.
In this way, after instrumentation, we can recover the original stack pointer through a pop.

In contrary, instrumentation using alternative stacks only needs to two instructions.
We save \code{\%rsp} to the scratch space so that later we can restore it,
and we change \code{\%rsp} to the start of the alternative stack.
Here, we do not need to align stack as we can pre-align the alternative stack when we allocate it.

Under our new alternative stack design,
unwinding in instrumentation code is the same as unwinding from the first instruction after the instrumentation.
In addition, instrumentation does not need to compensate the side effects caused by using the original stack.

\subsection{Viewing Instrumentation Costs}

\begin{figure}  
  \centering
  \includegraphics[page=1, width=0.4\textwidth]{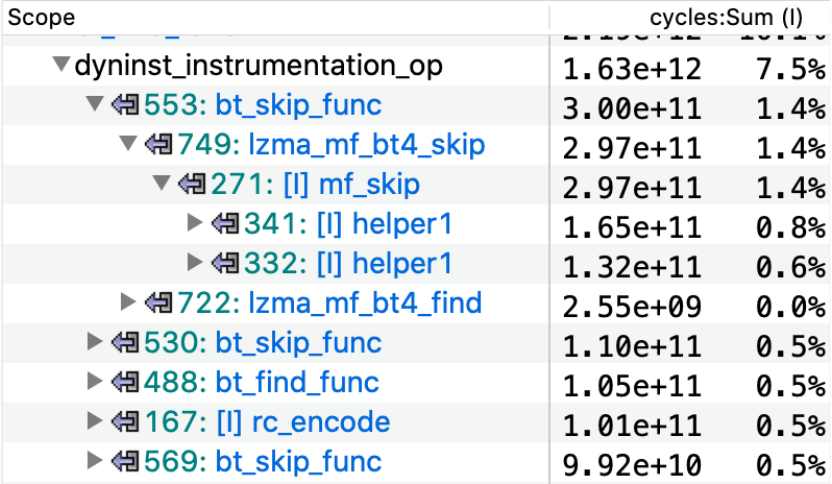}    
  \caption{Instrumentation costs incurred at different program locations shown in HPCViewer.}
  \label{fig:profile}
\end{figure}

\cref{fig:profile} shows using HPCViewer from the HPCToolkit to view instrumentation costs.
The scope column represents individual functions.
The \texttt{dyninst\_instrumentation\_op} represents a summary placeholder for instrumentation.
In this example, we can see that instrumentation costs 7.5\% of the total cycles.

Underneath the instrumentation placeholder, there are instrumentation costs attributed to different call paths.
Source line 553 in function \texttt{bt\_skip\_func} incurs 1.4\% instrumentation overhead,
while source line 530 incurs 0.5\% instrumentation overhead.

Binary rewriting application researchers can pinpoint the source of costs and design optimizations.

\section{Optimizing Binary Rewriting Applications}

We show to how use instrumentation profiling, static binary analysis,
and multi-version instrumentation to improve shadow stack and block coverage.

\subsection{Shadow Stack}

For shadow stack, our instrumentation requires a slot for the shadow stack pointer (\code{\%gs:0x0}),
and at most two slows for two scratch registers (\code{\%gs:0x8} and \code{\%gs:0x10}).
Memory locations after \code{\%gs:0x10} are used for the shadow stack.

The push and pop operations mirror regular call stack operations.
To support C++ exceptions and \code{setjmp}/\code{longjmp},
we repeatedly pop the values from the stack until a 
match or an under-flow occurs.
\Cref{fig:stack_push_basic} shows the instruction sequence for push.
The pop is similar (though with the loop).

As per \Cref{fig:stack_push_basic} the instruction sequence for a push consists of 9 instructions and 6
memory accesses. The pop operation incurs a similar overhead leading to about 20 instructions and 12 
memory accesses per function call from shadow stack instrumentation.
We use this instrumentation sequence as the baseline for shadow stack.

{
  \begin{listing}[t]
    \begin{minted}[escapeinside=||, mathescape=true]{gas}
    # Create scratch registers
    mov    %r10, %gs:0x8
    mov    %r11, %gs:0x10
    # Retrieve return address from the stack
    mov    0x0(%rsp),%r11
    # Retrieve the shadow stack pointer
    mov    %gs:0x0,%r10
    # Save return address to shadow stack
    mov    %r11,(%r10)
    # Move and save the shadow stack pointer
    lea    0x8(%r10),%r10
    mov    %r10,%gs:0x0
    # Restore scratch registers
    mov    %gs:0x8, %r10
    mov    %gs:0x10, %r11
    \end{minted}
    \caption{\small Instruction sequence for basic shadow stack push operation.}
    \label{fig:stack_push_basic}
    \end{listing}  
}

\subsubsection{Safe Function Elision}

We use static binary analysis to determine whether
a function is ``safe'' regarding return addresses
and elide instrumentation for safe functions.
Memory writes to global variables and 
stack variables within the current function frame (such as stack writes for spilling registers) are safe. 
If a function and all of its callees contain only safe memory writes, we do not instrument such function.
For indirect calls, as we cannot statically determine its targets;
we conservatively treat any function that may reach an indirect call as unsafe.

\subsubsection{Dead Register Chasing}

The basic shadow stack push and pop each need two GPRs as scratch space.
It is a well-known strategy to leverage register liveness analysis to identify dead registers.
We can avoid register save and restore if the scratch register is dead.

We observe that we can move our instrumentation to realize more dead registers
compared to fixing our instrumentation at function entry and exit.
For example, in~\Cref{fig:deadreg} if we instrument at the function entry,
\code{\%rbx} and \code{\%r10} are live because their values are then pushed to the stack.
If we move our shadow stack push operation after the push instruction sequence, we have two dead registers.
In this scenario, we need to adjust the offset for retrieving return address from 
the stack as we have pushed two GPRs to the stack.
We call this technique \textit{dead register chasing}.
Currently, we only move instrumentation within the same basic block
and we cannot move instrumentation beyond unsafe memory writes or unsafe function calls.

{
  \begin{listing}[t]
    \begin{minted}[escapeinside=||, mathescape=true]{gas}
    A: # No dead register at function entry
       push %rbx
       push %r10
       # %rbx and %r10 are dead here    
       mov 0x1234, %rbx
       mov %rdi, %r10
    \end{minted}
    \caption{\small An example showing the benefits of moving instrumentation for more dead registers.}
    \label{fig:deadreg}
    \end{listing}  
}

\subsubsection{Leaf Function Optimization}

Our profiling shows that in many cases, top hot functions are leaf functions.
Instead of storing the return address to the shadow stack,
we use a free register to implement a register stack frame.
\Cref{fig:stack_push_reg} shows an example of a register frame push operation.
Compared to the basic stack push operation shown in ~\Cref{fig:stack_push_basic}, 
a register frame push operation contains only 2 instructions.

We make two observations for the first instruction in ~\Cref{fig:stack_push_reg}.
We perform a register usage analysis of the leaf function to identify a free general 
purpose register (GPR). Note that we still need to save and restore the value of the chosen GPR 
because the compiler may not always emit ABI-compliant code (due to hand coded assembly etc.).
So, the unused register in the leaf function may carry live data for its caller.

{
  \begin{listing}[t]
    \begin{minted}[escapeinside=||, mathescape=true]{gas}
    # Save the register to the scratch space in
    # the shadow memory region
    mov    %r10,%gs:0x8
    # Retrieve return address
    mov    (%rsp),%r10    
    \end{minted}
    \caption{\small Instruction sequence for register frame push operation.}
    \label{fig:stack_push_reg}
    \end{listing}  
}

\subsubsection{Profile-Guided Inlining}

We use call path profiles to guide binary function inlining.
We mainly attribute instrumentation costs to a pair of the call site
and the corresponding callee,
and perform inlining from the most expensive call sites
until we cover 99\% of the instrumentation costs.

We handle two special cases.
First, call site attribution is not sufficient to determine whether cascaded inlined should be done.
Suppose we have three functions A, B, and C, where A calls B and B calls C.
In addition, the call site data tells us to inline C into B and inline B into A.
When we inline B into A, should we just inline the original B (without C) or inline the modified B (including C)?
If the call paths associated with costs of B calling C do not include A, then there is not need to inline C into A
as the the cost of B calling C is not incurred when A calls B;
otherwise, we should also inline C into A.

Second, it is possible that a call site and callee pair does not match.
Suppose A calls B, and B tail calls C.
When we take a sample in C, as B performs tail calls into C,
there is no stack frame or return addresses in B.
So, the call paths will look like that A directly calls C.
For such unmatched call sites and callees, we skip them.

\subsection{Block Coverage}

We use the block coverage implementation described by Khadra et al.~\cite{Khadra2020EBL} as the baseline.
We re-implement their approach with Dyninst to avoid
overhead differences caused by the binary rewriting tool.

\subsubsection{Thread Local Instrumentation Data}

The baseline implementation allocates static global memory to store coverage information.
The advantage is that instrumentation can be achieved with a single instruction
using PC-relative addressing: \code{movb 1, \%rip:off}. This instruction pattern is also used for other instrumentation
tasks, such as counting execution frequency of functions~\cite{WilliamsKing2020Egalito}.
We find that this instrumentation strategy will 
cause prohibitively overhead when running multi-threaded programs,
and the overhead will deteriorate with more threads.
We observed over 10X slowdown when using 8 threads.

The root cause was quickly revealed when we profile it.
We observed that instrumented code incurred a few times more 
last level cache misses than the uninstrumented code.
This points to the false sharing of the cache lines for the global memory used to store coverage information,
where different hardware threads may constantly write to the same cache line, causing this cache line to be
invalidated on other cores.

To mitigate this problem, we store coverage information in the alternative stack region described in \cref{sec:profile:as},
though we do not treat the memory region as a stack in this case.
A single instruction, \code{movb 1, \%gs:off}, can write coverage information to the alternative stack region.
We union the coverage information at the end of the execution to derive the coverage information for the process.

\subsubsection{Profile-Guided Instrumentation Data Allocation}

We then observed a performance problem with regard to the order of coverage data allocation.
In \cref{fig:datalayout}, 
we have two functions, whose function entries are instrumented.
We call the instrumentation at the entry of A (B) as $I_A$ ($I_B$).

Our profile shows that $I_A$ is a hot spot, which takes $1.0 \times 10^8$ cycles.
In addition, the call stacks of the samples for $I_A$ show that
all the cost by $I_A$ come from the call site at address 0xb6effd inside B.
Therefore, A should have the same execution frequency as B.
However, our profile shows that $I_B$ only takes $5.3 \times 10^7$ cycles, which is half of what $I_A$ takes.

Since $I_A$ and $I_B$ executed similar amount of time and both have the similar instruction (a memory write),
we concluded that $I_A$ is suffering from cache related problems.
It turned out that when we allocate which byte in the instrumentation data region to store coverage information for which basic block,
we (and existing work) just allocate one byte at time sequentially based on the processing order of the code,
which is typically in increasing order of the starting address of basic blocks.

In this case, $I_A$ writes coverage information for basic block at 0x416290 while
$I_B$ writes coverage information for basic block at 0xb64ff0.
So, the $I_A$ and $I_B$ are writing to memory location separated by many cache lines.
Each time when B calls A,
we will first access to one cache line for $I_B$ and then immediately access to another cache line for $I_A$.
So, there is no cache line reuse between $I_A$ and $I_B$.

{
  \begin{listing}[t]
    \begin{minted}[escapeinside=||, mathescape=true]{gas}
  A:
  # Instrumentation A takes 1.0 X 10^8 cycles 
  # Call stacks for all samples include B
  0x416290:  mov    0x8(%rdi),%rdx
  0x416294:  test   %rdx,%rdx

  B:
  # Instrumentation B takes 5.3 X 10^7 cycles
  0xb6eff0:  sub    $0x8,%rsp
  0xb6eff4:  mov    %edi,%esi
  0xb6eff6:  mov    0x61701b(%rip),%rdi
  0xb6effd:  callq  416290 <A>
  0xb6f002:  add    $0x8,%rsp
  0xb6f006:  retq       
    \end{minted}
    \caption{\small A performance problem in code coverage instrumentation revealed by performance analysis.}
    \label{fig:datalayout}
    \end{listing}  
}

To address this issue, we design a profile-guided coverage data allocation scheme.
We rank all function call pairs X and Y based on the costs attributed to function X calling Y.
From the highest pair, we try to allocate coverage data for X and Y as close as possible.
This way, high overhead pairs will enjoy better data locality.

\subsubsection{Profile-Guided Loop Cloning}

As described in \cref{sec:multi-version:loop},
idempotent instrumentation such as block coverage can benefit from 
loop cloning to elide instrumentation.
We attribute instrumentation costs to the basic blocks
and rank them in decreasing order of their costs.
We go through the ranked list and add a block to perform clone
if its belonging loop still contains fewer than $K$ blocks to clone.
Empirically, we set $K=5$, so we make at most $32$ copies for a loop.
We skip basic blocks that are not in any loop.

\subsubsection{Sampling Skid}

As each instrumentation point contains only one memory write instruction,
the sampling skid of performance events may seriously distort where the instrumentation costs come from.
Sampling skid refers to the delay between where a performance event happens the signal delivery to the performance tool.
For code coverage instrumentation, the skid can cause instrumentation costs to be attributed to original code.
In some cases, we have seen all costs incurred by instrumentation are attributed to original code,
which defeats the purpose of profiling analysis of instrumentation and profile-guided instrumentation.

Vendors provide hardware support to mitigate sampling skid,
including Intel's precise event-based sampling (PEBS)
and AMD's instruction-based sampling (IBS).
However, recent studies have shown that these ``precise'' sampling technologies may still be imprecise~\cite{Xu2019CWT, Yi2020PEBS}.

In our case, profiles collected using Intel's PEBS for block coverage enabled
us to identify the data allocation problem described above, and are effective for performing profile-guided instrumentation.
It is an interesting future research topic to see whether a sophisticated handling of sampling skid can yield better results.

\section{Evaluation}

We evaluate our implementation with SPEC CPU 2017 and two real world software.

\subsection{SPEC CPU 2017}

We are able to compile and run 19 of the 20 available benchmarks.
627.cam4\_s does not compile on any of the systems, so is excluded.
8 of the 19 benchmarks are written in Fortran or have Fortran components.
The other benchmarks are in C/C++.

All profile-guided versions utilize the training workloads in SPEC CPU 2017,
which are different from the reference workloads that are used for reporting the results.
Some benchmarks provide multiple training workloads.
For these benchmarks, the final profiles are the aggregation of the individual profiles.

We run all instrumentation versions 10 times using 8 hardware threads 
and use the selected runs by SPEC CPU 2017 to compute overhead.
\cref{table:spec2017:results} shows the result summary.
The \texttt{empty} shows the overhead of instrumenting every basic block with empty instrumentation
and represents the tool overhead caused by incremental CFG patching~\cite{Meng2021ICFGP}.

{
\begin{table}
  \centering
  \caption{Shadow stack and block coverage results.}
  \begin{tabular}{|l | r  | r | r | r | }
  \hline    
              &  \multicolumn{2}{c|}{Time overhead} & \multicolumn{2}{c|}{Size increase} \\  
  \cline{2-5}                         
              &   max       &        mean      &    max       &     mean   \\                
  \hline
  \texttt{empty}       &  1.0\%      &       0.20\%     &   101.53\%   &    64.27\%  \\
  \hline
  \hline
  \texttt{SS-base}     &  31.0\%     &     7.6\%        &   141.4\%    &   81.0\%   \\
  \texttt{SS-static}   &  13.3\%     &     3.1\%        &   119.0\%    &   73.8\%   \\
  \texttt{SS-inline}   &  10.0\%     &     1.4\%        &   126.2\%    &   78.5\%   \\
  \hline
  \hline
  \texttt{BC-base}     &  1495.4\%   &  161.3\%         &   114.6\%    &   76.4\%   \\
  \texttt{BC-thread}   &  51.1\%     &    9.8\%         &   126.5\%    &   77.7\%   \\
  \texttt{BC-loop}     &  14.0\%     &    5.9\%         &  1813.8\%    &  145.0\%   \\
  \texttt{BC-data}     &  12.1\%     &    4.0\%         &  1813.8\%    &  145.0\%   \\
  \hline
  \end{tabular}
  \label{table:spec2017:results}
\end{table}
}

\subsubsection{Shadow Stack}

We compare three versions:
\texttt{SS-base} represents a baseline binary level shadow stack implementation,
where we instrument every function entry and exit with the baseline instrumentation sequence shown in \cref{fig:stack_push_basic};
\texttt{SS-static} represents an improved shadow stack implementation using static binary analysis
to elide instrumentation of safe functions, leaf function frame optimization, and dead register chasing;
\texttt{SS-inline} is \texttt{SS-static} with profile-guided binary function inlining.
The experiments for shadow stack are done on an AMD EPYC 7402 server,
which has 512GB memory and runs Red Hat 8.3.
We use the system default compiler gcc-8.3.1.
Profile data is collected with CPU timers.

Overall, we can see that \texttt{SS-static} significantly reduces runtime overhead compared to \texttt{SS-base},
showing that static binary analysis and 
careful instrumentation instruction sequence designs are two power weapons to optimize
binary instrumentation applications.
Equipped with call path profiling, users can identify to where to pay attention.
\texttt{SS-inline} cut the average overhead by half compared to \texttt{SS-static},
showing that profile-guided inlining is a useful for optimizing binary instrumentation.

Next, we dive into the results of three benchmarks to better understand the capability of profile-guided inlining.
One of the most successful benchmark for \texttt{SS-inline} is 620.omnetpp\_s, which is written in C++.
Our profile shows that an indirect function call, which implements a virtual function call, incurred the most costs.
And almost all costs of this indirect call site goes to a single function.
The combination of indirect call promotion and function inlining can eliminate most of the costs.
It is often challenging for static binary analysis to analyze indirect calls.
\texttt{SS-static} incurred 10.6\% overhead for this benchmark and
its overhead was reduced to 3.9\% with \texttt{SS-inline}.

\texttt{SS-inline} incurred the highest overhead with 600.perlbench\_s.
Its overhead is centered around an indirect call that serves as a dispatcher
to call different interpreter operations. This indirect call can go to over 60 different functions.
It returns the next operation to invoke and is executed inside a small loop repeatedly that fits in a single cache line.
We observed that when attempting to promote more than 5 indirect call targets for this indirect call,
instruction cache pressure will make performance worse than without any inlining.

The call site profile for 602.gcc\_s exhibits a long-tail distribution,
which means that we have to inline much more code for this benchmark to cover instruction
We find that the net effect of binary function inlining is negligible compared without inlining.

\subsubsection{Block Coverage}

For block coverage, we compare four versions:
\texttt{BC-base} represents a baseline of binary-level block coverage implementation~\cite{Khadra2020EBL};
\texttt{BC-thread} represents the version that store coverage information in thread local memory;
\texttt{BC-loop} represents \texttt{BC-thread} with profile-guided multi-version loop instrumentation;
\texttt{BC-data} is \texttt{BC-thread} with profile-guided coverage data allocation.
The experiments are done on an Intel(R) Xeon(R) CPU E5-2695 server,
which has 128GB memory and runs Red Hat 7.9.
We use a gcc-7.3.0 built by the system admin as the system compiler for Red Hat 7.9 is gcc-4.8.5, which is too old.
Profile data is collected with the PEBS cycle counter to mitigate the sampling skid.

\texttt{BC-base} suffers from false sharing, incurring over 100\% overhead on average.
\texttt{BC-thread} stores coverage information in thread local memory regions, thus
avoiding false sharing. \texttt{BC-base} and \texttt{BC-thread} have similar overhead 
numbers for benchmarks that do not use multi-threading.
\texttt{BC-thread} incurs slightly higher size overhead as the instruction that writes
a memory location related to \code{\%gs} is two-byte longer than the instruction that
is PC-relative.

\texttt{BC-thread} incurred highest overhead, 51.1\%, with 644.nab\_s.
\texttt{BC-loop} and \texttt{BC-data} reduced the overhead to 10.0\% and 3.7\%, respectively.

\texttt{BC-loop} and \texttt{BC-data} have the same size overhead as \texttt{BC-data} only reorders
instrumentation data allocation.
Both incurred over 18X binary size increase for 648.exchange2\_s.
This dramatic size increase is caused by making copies of loops that are over a few KB in size.
The net effect is that \texttt{BC-loop} does not reduce overhead for this benchmark compared to \texttt{BC-thread}

\subsection{Real World Software}

We also evaluated shadow stack and block coverage instrumentation using two real world software:
Apache HTTP Server and Firefox. The experiments were done on the Intel system mentioned above.

\subsubsection{Apache HTTP Server}
We compiled httpd-2.4.43 and 
instrumented the server executable \code{httpd}.
We used the provided benchmarking program \code{ab}
to measure instrumentation overhead by transferring a 595KB HTML file 100,000 times.
We collect call path profiles by transferring the same file, but only 10,000 times.
The experiments were repeated 20 times.

For shadow stack instrumentation,
\texttt{SS-base} caused 18.4\% throughput reduction and 22.3\% latency increase;
\texttt{SS-static} improved these numbers to
10.7\% throughput reduction and 12.0\% latency increase;
\texttt{SS-inline} performed the best, causing
3.3\% throughput reduction and 3.6\% latency increase.

For block coverage instrumentation, all versions incurred negligible overhead.

\subsubsection{Firefox} 
We used the Firefox shipped with Red Hat 7.9, which is 78.4.0.
We instrument the \code{libxul.so} library in Firefox.
This library is the main component of Firefox and its \code{.text} section is over 100MB in size.

To our surprise, none of our profile-guided instrumentation for shadow stack or block coverage
can reduce instrumentation overhead, even when we collected profiles using the same benchmarks for testing.

The idea of multi-version binary rewriting centers on creating multiple copies of code to elide instrumentation,
which increase the code size.
If the root cause of binary instrumentation overhead are data accesses,
multi-version binary rewriting can be beneficial.
However, if the root cause of instrumentation overhead is related to instruction fetching,
multi-version binary rewriting will not be helpful.

For Firefox, when we replaced the actual instrumentation with nops in the same length, we found
that the nop instrumentation incurred similar overhead as the actual instrumentation.
The nop instrumentation causes no additional data accesses but has the effect on instruction cache
as the original instrumentation. This confirms that the instrumentation overhead for Firefox is instruction
fetching related.

We also found that for \code{libxu.so}, even after dominator analysis, we need to instrument over 2 million basic blocks.
Each memory write instruction in \texttt{BC-thread} is 9-byte long.
So the instrumentation increased the code size by over 18MB!
The profiles for Firefox also exhibited long-tail distributions.
These two factors together explains why instrumentation overhead for Firefox is mainly caused by instruction fetching.

While it is disappointing, we believe this ``negative'' result and our analysis above are
beneficial for understanding the scenarios where our new techniques can be useful.

\section{Discussion}

\textbf{Last Branch Record:}
Recent studies on PGO~\cite{Panchenko2019BOLT, Panchenko2021LBOLT, Duta2021PIBE} utilize the last branch records (LBR)
provided by Intel to extract execution counts of control flow edges.
LBR has very limited availability on non-Intel hardware.
Our work enables call path profiling for binary instrumentation,
which can collect performance information utilizing timer interrupts and hardware performance counters,
thus has a wider applicability.
We note that our approach can also be improved by incorporating LBR when available.
We mentioned that call path profiling will miss functions performing tail calls,
which can potentially be captured by LBR.

\textbf{Application Specific Issues:}
From binary rewriting application researchers' perspectives,
there are other application specific issues to consider besides
the instrumentation overhead.
For shadow stack, the shadow memory region must be isolated from the original program,
which is an important but orthogonal research area~\cite{Sehr2010ASF, Burow2018CFIXXOT}.
There are also concerns on cross-thread attacks~\cite{Xu2019CONFIRM}.
For code coverage, edge coverage may reveal more information compared to block coverage~\cite{Zhu2020CSIFuzz}.
Our message is that by showing we can improve two typical binary rewriting applications in conventional settings,
our can benefit binary rewriting application researchers in other scenarios.

\section{Conclusion}

We have presented multi-version binary rewriting,
a new technique for optimizing static binary instrumentation.
We enabled call path profiling for static binary instrumentation,
designed primitive binary transformation operations to support complex binary transformation, 
and applied instrumentation performance analysis and profile-guided binary transformation to optimize
shadow stack and block coverage, two typical binary rewriting applications.
Our results of SPEC CPU 2017 benchmarks, Apache HTTPD Server, and Firefox
show that our techniques work well on programs where the root cause of instrumentation overhead is data accesses,
and point out future research direction for designing optimization strategies to 
reduce instrumentation overhead caused by instruction fetching issues.


\bibliographystyle{ACM-Reference-Format}
\bibliography{rewriting}


\end{document}

%% file: config.tex
\usepackage{tikz}
\usepackage{amsmath}
\usepackage{amsfonts}
\usepackage{xcolor, colortbl}
\usepackage{collcell}
\usetikzlibrary{positioning}
\usetikzlibrary{arrows}
\usetikzlibrary{plotmarks}
\usepackage[frozencache=true,cachedir=./minted_cache]{minted}
\usepackage{pgfplots}
\usepackage{pgfplotstable}
\pgfplotsset{compat=1.15}
\usepackage{tabularx}
\usepackage{diagbox}
\usepackage{multirow}
\usepackage{caption}
\usepackage{subcaption}
\usepackage{enumitem}
\usepackage{bigdelim}
\usepackage{calc}
\usepackage{fancyhdr}
\usepackage{hyperref}
\usepackage{algorithm}
\usepackage[noabbrev,capitalize]{cleveref}
\usepackage{diagbox}
\usepackage{graphicx}
\usepackage{bm}
\usepackage{pifont}
\usepackage{adjustbox}
\usepackage{diagbox}
\usepackage{algpseudocode}
\usepackage{mathtools}

\crefrangelabelformat{Figure}{#3#1#4--#5\crefstripprefix{#1}{#2}#6}

\newcommand{\code}[1]{{\texttt{#1}}}

\newlength\figureheight
\newlength\figurewidth
\setlist{topsep=0pt,nolistsep}

\usepackage{listings}

%% file: versioned-cfg-editing-paper.bbl

\begin{thebibliography}{45}


\ifx \showCODEN    \undefined \def \showCODEN     #1{\unskip}     \fi
\ifx \showDOI      \undefined \def \showDOI       #1{#1}\fi
\ifx \showISBNx    \undefined \def \showISBNx     #1{\unskip}     \fi
\ifx \showISBNxiii \undefined \def \showISBNxiii  #1{\unskip}     \fi
\ifx \showISSN     \undefined \def \showISSN      #1{\unskip}     \fi
\ifx \showLCCN     \undefined \def \showLCCN      #1{\unskip}     \fi
\ifx \shownote     \undefined \def \shownote      #1{#1}          \fi
\ifx \showarticletitle \undefined \def \showarticletitle #1{#1}   \fi
\ifx \showURL      \undefined \def \showURL       {\relax}        \fi
\providecommand\bibfield[2]{#2}
\providecommand\bibinfo[2]{#2}
\providecommand\natexlab[1]{#1}
\providecommand\showeprint[2][]{arXiv:#2}

\bibitem[\protect\citeauthoryear{Adhianto, Banerjee, Fagan, Krentel, Marin,
  Mellor-Crummey, and Tallent}{Adhianto et~al\mbox{.}}{2010}]%
        {HPCToolkit2010Adhianto}
\bibfield{author}{\bibinfo{person}{L. Adhianto}, \bibinfo{person}{S. Banerjee},
  \bibinfo{person}{M. Fagan}, \bibinfo{person}{M. Krentel}, \bibinfo{person}{G.
  Marin}, \bibinfo{person}{J. Mellor-Crummey}, {and} \bibinfo{person}{N.~R.
  Tallent}.} \bibinfo{year}{2010}\natexlab{}.
\newblock \showarticletitle{HPCTOOLKIT: Tools for Performance Analysis of
  Optimized Parallel Programs Http://Hpctoolkit.Org}.
\newblock \bibinfo{journal}{\emph{Concurrency and Computation: Practice \&
  Experience}} \bibinfo{volume}{22}, \bibinfo{number}{6} (\bibinfo{date}{Apr.}
  \bibinfo{year}{2010}), \bibinfo{pages}{685–701}.
\newblock
\showISSN{1532-0626}


\bibitem[\protect\citeauthoryear{Agrawal}{Agrawal}{1994}]%
        {Agrawal1994Coverage}
\bibfield{author}{\bibinfo{person}{Hiralal Agrawal}.}
  \bibinfo{year}{1994}\natexlab{}.
\newblock \showarticletitle{Dominators, Super Blocks, and Program Coverage}. In
  \bibinfo{booktitle}{\emph{Proceedings of the 21st ACM SIGPLAN-SIGACT
  Symposium on Principles of Programming Languages}}
  \emph{(\bibinfo{series}{POPL '94})}. \bibinfo{address}{Portland, Oregon,
  USA}, \bibinfo{pages}{25–34}.
\newblock
\showISBNx{0897916360}
\urldef\tempurl%
\url{https://doi.org/10.1145/174675.175935}
\showDOI{\tempurl}


\bibitem[\protect\citeauthoryear{Arnold, Fink, Sarkar, and Sweeney}{Arnold
  et~al\mbox{.}}{2000}]%
        {Arnold2000ACS}
\bibfield{author}{\bibinfo{person}{Matthew Arnold}, \bibinfo{person}{Stephen
  Fink}, \bibinfo{person}{Vivek Sarkar}, {and} \bibinfo{person}{Peter~F.
  Sweeney}.} \bibinfo{year}{2000}\natexlab{}.
\newblock \showarticletitle{A Comparative Study of Static and Profile-Based
  Heuristics for Inlining}. In \bibinfo{booktitle}{\emph{Proceedings of the ACM
  SIGPLAN Workshop on Dynamic and Adaptive Compilation and Optimization}}
  \emph{(\bibinfo{series}{DYNAMO '00})}. \bibinfo{pages}{52–64}.
\newblock
\showISBNx{1581132417}
\urldef\tempurl%
\url{https://doi.org/10.1145/351397.351416}
\showURL{%
\tempurl}


\bibitem[\protect\citeauthoryear{Ayers, Schooler, and Gottlieb}{Ayers
  et~al\mbox{.}}{1997}]%
        {Ayers1997AI}
\bibfield{author}{\bibinfo{person}{Andrew Ayers}, \bibinfo{person}{Richard
  Schooler}, {and} \bibinfo{person}{Robert Gottlieb}.}
  \bibinfo{year}{1997}\natexlab{}.
\newblock \showarticletitle{Aggressive Inlining}. In
  \bibinfo{booktitle}{\emph{Proceedings of the ACM SIGPLAN 1997 Conference on
  Programming Language Design and Implementation}} (Las Vegas, Nevada, USA)
  \emph{(\bibinfo{series}{PLDI '97})}. \bibinfo{pages}{134–145}.
\newblock
\showISBNx{0897919076}
\urldef\tempurl%
\url{https://doi.org/10.1145/258915.258928}
\showURL{%
\tempurl}


\bibitem[\protect\citeauthoryear{Ben~Khadra, Stoffel, and Kunz}{Ben~Khadra
  et~al\mbox{.}}{2020}]%
        {Khadra2020EBL}
\bibfield{author}{\bibinfo{person}{M.~Ammar Ben~Khadra},
  \bibinfo{person}{Dominik Stoffel}, {and} \bibinfo{person}{Wolfgang Kunz}.}
  \bibinfo{year}{2020}\natexlab{}.
\newblock \showarticletitle{Efficient Binary-Level Coverage Analysis}. In
  \bibinfo{booktitle}{\emph{Proceedings of the 28th ACM Joint Meeting on
  European Software Engineering Conference and Symposium on the Foundations of
  Software Engineering (ESEC/FSE)}} (Virtual Event, USA).
  \bibinfo{pages}{1153–1164}.
\newblock
\showISBNx{9781450370431}
\urldef\tempurl%
\url{https://doi.org/10.1145/3368089.3409694}
\showURL{%
\tempurl}


\bibitem[\protect\citeauthoryear{Bernat and Miller}{Bernat and Miller}{2011}]%
        {Bernat2011AWAT}
\bibfield{author}{\bibinfo{person}{Andrew~R. Bernat} {and}
  \bibinfo{person}{Barton~P. Miller}.} \bibinfo{year}{2011}\natexlab{}.
\newblock \showarticletitle{Anywhere, Any-Time Binary Instrumentation}. In
  \bibinfo{booktitle}{\emph{10th ACM SIGPLAN-SIGSOFT Workshop on Program
  Analysis for Software Tools}} \emph{(\bibinfo{series}{PASTE'11})}.
  \bibinfo{address}{Szeged, Hungary}, \bibinfo{pages}{9–16}.
\newblock


\bibitem[\protect\citeauthoryear{Bernat and Miller}{Bernat and Miller}{2012}]%
        {Bernat2012SBE}
\bibfield{author}{\bibinfo{person}{Andrew~R. Bernat} {and}
  \bibinfo{person}{Barton~P. Miller}.} \bibinfo{year}{2012}\natexlab{}.
\newblock \showarticletitle{Structured Binary Editing with a CFG Transformation
  Algebra}. In \bibinfo{booktitle}{\emph{2012 19th Working Conference on
  Reverse Engineering (WCRE)}}. \bibinfo{address}{Kingston, ON, Canada},
  \bibinfo{pages}{9–18}.
\newblock


\bibitem[\protect\citeauthoryear{Bernat, Roundy, and Miller}{Bernat
  et~al\mbox{.}}{2011}]%
        {Bernat2011ESR}
\bibfield{author}{\bibinfo{person}{Andrew~R. Bernat}, \bibinfo{person}{Kevin~A.
  Roundy}, {and} \bibinfo{person}{Barton~P. Miller}.}
  \bibinfo{year}{2011}\natexlab{}.
\newblock \showarticletitle{Efficient, Sensitivity Resistant Binary
  Instrumentation}. In \bibinfo{booktitle}{\emph{The International Symposium on
  Software Testing and Analysis (ISSTA)}}. \bibinfo{address}{Toronto, Canada}.
\newblock


\bibitem[\protect\citeauthoryear{Burow, McKee, Carr, and Payer}{Burow
  et~al\mbox{.}}{2018}]%
        {Burow2018CFIXXOT}
\bibfield{author}{\bibinfo{person}{Nathan Burow}, \bibinfo{person}{Derrick
  McKee}, \bibinfo{person}{Scott~A. Carr}, {and} \bibinfo{person}{Mathias
  Payer}.} \bibinfo{year}{2018}\natexlab{}.
\newblock \showarticletitle{CFIXX: Object Type Integrity for C++}. In
  \bibinfo{booktitle}{\emph{NDSS}}.
\newblock


\bibitem[\protect\citeauthoryear{Chang and Hwu}{Chang and Hwu}{1989}]%
        {Chang1989IFE}
\bibfield{author}{\bibinfo{person}{P.~P. Chang} {and} \bibinfo{person}{W.-W.
  Hwu}.} \bibinfo{year}{1989}\natexlab{}.
\newblock \showarticletitle{Inline Function Expansion for Compiling C
  Programs}.
\newblock \bibinfo{journal}{\emph{ACM SIGPLAN Notice}} \bibinfo{volume}{24},
  \bibinfo{number}{7} (\bibinfo{date}{June} \bibinfo{year}{1989}),
  \bibinfo{pages}{246–257}.
\newblock
\showISSN{0362-1340}
\urldef\tempurl%
\url{https://doi.org/10.1145/74818.74840}
\showURL{%
\tempurl}


\bibitem[\protect\citeauthoryear{Chen, Li, and Moseley}{Chen
  et~al\mbox{.}}{2016}]%
        {Chen2016AutoFDO}
\bibfield{author}{\bibinfo{person}{Dehao Chen}, \bibinfo{person}{David~Xinliang
  Li}, {and} \bibinfo{person}{Tipp Moseley}.} \bibinfo{year}{2016}\natexlab{}.
\newblock \showarticletitle{AutoFDO: Automatic Feedback-directed Optimization
  for Warehouse-scale Applications}. In \bibinfo{booktitle}{\emph{2016
  International Symposium on Code Generation and Optimization (CGO)}}.
  \bibinfo{address}{Barcelona, Spain}.
\newblock


\bibitem[\protect\citeauthoryear{{Chen}, {Vachharajani}, {Hundt}, {Li},
  {Eranian}, {Chen}, and {Zheng}}{{Chen} et~al\mbox{.}}{2013}]%
        {Chen2013THE}
\bibfield{author}{\bibinfo{person}{D. {Chen}}, \bibinfo{person}{N.
  {Vachharajani}}, \bibinfo{person}{R. {Hundt}}, \bibinfo{person}{X. {Li}},
  \bibinfo{person}{S. {Eranian}}, \bibinfo{person}{W. {Chen}}, {and}
  \bibinfo{person}{W. {Zheng}}.} \bibinfo{year}{2013}\natexlab{}.
\newblock \showarticletitle{Taming Hardware Event Samples for Precise and
  Versatile Feedback Directed Optimizations}.
\newblock \bibinfo{journal}{\emph{IEEE Trans. Comput.}} \bibinfo{volume}{62},
  \bibinfo{number}{2} (\bibinfo{year}{2013}), \bibinfo{pages}{376--389}.
\newblock
\urldef\tempurl%
\url{https://doi.org/10.1109/TC.2011.233}
\showDOI{\tempurl}


\bibitem[\protect\citeauthoryear{Cohn, Goodwin, and Lowney}{Cohn
  et~al\mbox{.}}{1998}]%
        {Cohn1998OAE}
\bibfield{author}{\bibinfo{person}{Robert~S. Cohn}, \bibinfo{person}{David~W.
  Goodwin}, {and} \bibinfo{person}{P.~Geoffrey Lowney}.}
  \bibinfo{year}{1998}\natexlab{}.
\newblock \showarticletitle{Optimizing Alpha Executables on Windows NT with
  Spike}.
\newblock \bibinfo{journal}{\emph{Digital Technical Journal}}
  \bibinfo{volume}{9}, \bibinfo{number}{4} (\bibinfo{date}{April}
  \bibinfo{year}{1998}), \bibinfo{pages}{3–20}.
\newblock
\showISSN{0898-901X}


\bibitem[\protect\citeauthoryear{Dang, Maniatis, and Wagner}{Dang
  et~al\mbox{.}}{2015}]%
        {dang2015performance}
\bibfield{author}{\bibinfo{person}{Thurston~HY Dang}, \bibinfo{person}{Petros
  Maniatis}, {and} \bibinfo{person}{David Wagner}.}
  \bibinfo{year}{2015}\natexlab{}.
\newblock \showarticletitle{The performance cost of shadow stacks and stack
  canaries}. In \bibinfo{booktitle}{\emph{Proceedings of the 10th ACM Symposium
  on Information, Computer and Communications Security}}. ACM,
  \bibinfo{pages}{555--566}.
\newblock


\bibitem[\protect\citeauthoryear{Di~Federico, Payer, and Agosta}{Di~Federico
  et~al\mbox{.}}{2017}]%
        {DiFederico2017RUB}
\bibfield{author}{\bibinfo{person}{Alessandro Di~Federico},
  \bibinfo{person}{Mathias Payer}, {and} \bibinfo{person}{Giovanni Agosta}.}
  \bibinfo{year}{2017}\natexlab{}.
\newblock \showarticletitle{Rev.Ng: A Unified Binary Analysis Framework to
  Recover CFGs and Function Boundaries}. In \bibinfo{booktitle}{\emph{26th
  International Conference on Compiler Construction (CC)}}.
  \bibinfo{address}{Austin, TX, USA}.
\newblock


\bibitem[\protect\citeauthoryear{Dinesh, Burow, Xu, and Payer}{Dinesh
  et~al\mbox{.}}{2020}]%
        {Dinesh2020RetroWrite}
\bibfield{author}{\bibinfo{person}{Sushant Dinesh}, \bibinfo{person}{Nathan
  Burow}, \bibinfo{person}{Dongyan Xu}, {and} \bibinfo{person}{Mathias Payer}.}
  \bibinfo{year}{2020}\natexlab{}.
\newblock \showarticletitle{RetroWrite: Statically Instrumenting COTS Binaries
  for Fuzzing and Sanitization}. In \bibinfo{booktitle}{\emph{41st IEEE
  Symposium on Security and Privacy (Oakland)}}.
\newblock


\bibitem[\protect\citeauthoryear{Duck, Gao, and Roychoudhury}{Duck
  et~al\mbox{.}}{2020}]%
        {Duck2020BRW}
\bibfield{author}{\bibinfo{person}{Gregory~J. Duck}, \bibinfo{person}{Xiang
  Gao}, {and} \bibinfo{person}{Abhik Roychoudhury}.}
  \bibinfo{year}{2020}\natexlab{}.
\newblock \showarticletitle{Binary Rewriting without Control Flow Recovery}. In
  \bibinfo{booktitle}{\emph{41st ACM SIGPLAN Conference on Programming Language
  Design and Implementation (PLDI)}}. \bibinfo{address}{London, UK}.
\newblock


\bibitem[\protect\citeauthoryear{Duta, Giuffrida, Bos, and van~der Kouwe}{Duta
  et~al\mbox{.}}{2021}]%
        {Duta2021PIBE}
\bibfield{author}{\bibinfo{person}{Victor Duta}, \bibinfo{person}{Cristiano
  Giuffrida}, \bibinfo{person}{Herbert Bos}, {and} \bibinfo{person}{Erik
  van~der Kouwe}.} \bibinfo{year}{2021}\natexlab{}.
\newblock \showarticletitle{PIBE: Practical Kernel Control-Flow Hardening with
  Profile-Guided Indirect Branch Elimination}. In
  \bibinfo{booktitle}{\emph{Proceedings of the 26th ACM International
  Conference on Architectural Support for Programming Languages and Operating
  Systems}} (Virtual, USA) \emph{(\bibinfo{series}{ASPLOS 2021})}.
  \bibinfo{pages}{743–757}.
\newblock
\urldef\tempurl%
\url{https://doi.org/10.1145/3445814.3446740}
\showDOI{\tempurl}


\bibitem[\protect\citeauthoryear{Flores-Montoya and Schulte}{Flores-Montoya and
  Schulte}{2020}]%
        {Montoya2020DD}
\bibfield{author}{\bibinfo{person}{Antonio Flores-Montoya} {and}
  \bibinfo{person}{Eric Schulte}.} \bibinfo{year}{2020}\natexlab{}.
\newblock \showarticletitle{Datalog Disassembly}. In
  \bibinfo{booktitle}{\emph{29th {USENIX} Security Symposium ({USENIX} Security
  20)}}. \bibinfo{pages}{1075--1092}.
\newblock
\showISBNx{978-1-939133-17-5}
\urldef\tempurl%
\url{https://www.usenix.org/conference/usenixsecurity20/presentation/flores-montoya}
\showURL{%
\tempurl}


\bibitem[\protect\citeauthoryear{{Google}}{{Google}}{2021}]%
        {PROPELLER}
\bibfield{author}{\bibinfo{person}{{Google}}.} \bibinfo{year}{accessed April
  24, 2021}\natexlab{}.
\newblock \bibinfo{title}{{PROPELLER: Profile Guided Optimizing Large Scale
  LLVM-based Relinker}, \url{https://github.com/google/llvm-propeller}}.
\newblock
\newblock


\bibitem[\protect\citeauthoryear{Gu and Mellor-Crummey}{Gu and
  Mellor-Crummey}{2018}]%
        {Gu2018DDR}
\bibfield{author}{\bibinfo{person}{Yizi Gu} {and} \bibinfo{person}{John
  Mellor-Crummey}.} \bibinfo{year}{2018}\natexlab{}.
\newblock \showarticletitle{Dynamic Data Race Detection for OpenMP Programs}.
  In \bibinfo{booktitle}{\emph{International Conference for High Performance
  Computing, Networking, Storage, and Analysis (SC)}}.
  \bibinfo{address}{Dallas, Texas}.
\newblock


\bibitem[\protect\citeauthoryear{Hall}{Hall}{1992}]%
        {Hall1992CPP}
\bibfield{author}{\bibinfo{person}{Robert~J. Hall}.}
  \bibinfo{year}{1992}\natexlab{}.
\newblock \showarticletitle{Call Path Profiling}. In
  \bibinfo{booktitle}{\emph{Proceedings of the 14th International Conference on
  Software Engineering}} (Melbourne, Australia) \emph{(\bibinfo{series}{ICSE
  '92})}. \bibinfo{pages}{296–306}.
\newblock
\showISBNx{0897915046}
\urldef\tempurl%
\url{https://doi.org/10.1145/143062.143147}
\showDOI{\tempurl}


\bibitem[\protect\citeauthoryear{Hawkins, Hiser, Co, Nguyen-Tuong, and
  Davidson}{Hawkins et~al\mbox{.}}{2017}]%
        {Hawkins2017Zipr}
\bibfield{author}{\bibinfo{person}{William~H. Hawkins},
  \bibinfo{person}{Jason~D. Hiser}, \bibinfo{person}{Michele Co},
  \bibinfo{person}{Anh Nguyen-Tuong}, {and} \bibinfo{person}{Jack~W.
  Davidson}.} \bibinfo{year}{2017}\natexlab{}.
\newblock \showarticletitle{Zipr: Efficient Static Binary Rewriting for
  Security}. In \bibinfo{booktitle}{\emph{2017 47th Annual IEEE/IFIP
  International Conference on Dependable Systems and Networks (DSN)}}.
  \bibinfo{address}{Denver, CO, USA}, \bibinfo{pages}{559--566}.
\newblock
\urldef\tempurl%
\url{https://doi.org/10.1109/DSN.2017.27}
\showDOI{\tempurl}


\bibitem[\protect\citeauthoryear{Meng and Liu}{Meng and Liu}{2021}]%
        {Meng2021ICFGP}
\bibfield{author}{\bibinfo{person}{Xiaozhu Meng} {and} \bibinfo{person}{Weijie
  Liu}.} \bibinfo{year}{2021}\natexlab{}.
\newblock \showarticletitle{Incremental CFG Patching for Binary Rewriting}. In
  \bibinfo{booktitle}{\emph{Proceedings of the 26th ACM International
  Conference on Architectural Support for Programming Languages and Operating
  Systems (ASPLOS)}} (Virtual Event).
\newblock


\bibitem[\protect\citeauthoryear{Meng and Miller}{Meng and Miller}{2016}]%
        {Meng2016BinaryNotEasy}
\bibfield{author}{\bibinfo{person}{Xiaozhu Meng} {and}
  \bibinfo{person}{Barton~P. Miller}.} \bibinfo{year}{2016}\natexlab{}.
\newblock \showarticletitle{Binary Code Is Not Easy}. In
  \bibinfo{booktitle}{\emph{The International Symposium on Software Testing and
  Analysis (ISSTA)}}. \bibinfo{address}{Saarbr\"{u}cken, Germany}.
\newblock


\bibitem[\protect\citeauthoryear{Ottoni and Maher}{Ottoni and Maher}{2017}]%
        {Ottoni2017OFP}
\bibfield{author}{\bibinfo{person}{Guilherme Ottoni} {and}
  \bibinfo{person}{Bertrand Maher}.} \bibinfo{year}{2017}\natexlab{}.
\newblock \showarticletitle{Optimizing Function Placement for Large-Scale
  Data-Center Applications}. In \bibinfo{booktitle}{\emph{Proceedings of the
  2017 International Symposium on Code Generation and Optimization (CGO)}}.
  \bibinfo{address}{Austin, USA}.
\newblock


\bibitem[\protect\citeauthoryear{Panchenko, Auler, Nell, and Ottoni}{Panchenko
  et~al\mbox{.}}{2019}]%
        {Panchenko2019BOLT}
\bibfield{author}{\bibinfo{person}{Maksim Panchenko}, \bibinfo{person}{Rafael
  Auler}, \bibinfo{person}{Bill Nell}, {and} \bibinfo{person}{Guilherme
  Ottoni}.} \bibinfo{year}{2019}\natexlab{}.
\newblock \showarticletitle{BOLT: A Practical Binary Optimizer for Data Centers
  and Beyond}. In \bibinfo{booktitle}{\emph{Proceedings of the 2019 IEEE/ACM
  International Symposium on Code Generation and Optimization (CGO)}}.
  \bibinfo{address}{Washington, DC, USA}, \bibinfo{pages}{2–14}.
\newblock


\bibitem[\protect\citeauthoryear{Panchenko, Auler, Sakka, and Ottoni}{Panchenko
  et~al\mbox{.}}{2021}]%
        {Panchenko2021LBOLT}
\bibfield{author}{\bibinfo{person}{Maksim Panchenko}, \bibinfo{person}{Rafael
  Auler}, \bibinfo{person}{Laith Sakka}, {and} \bibinfo{person}{Guilherme
  Ottoni}.} \bibinfo{year}{2021}\natexlab{}.
\newblock \showarticletitle{Lightning BOLT: Powerful, Fast, and Scalable Binary
  Optimization}. In \bibinfo{booktitle}{\emph{Proceedings of the 30th ACM
  SIGPLAN International Conference on Compiler Construction}} (Virtual,
  Republic of Korea) \emph{(\bibinfo{series}{CC 2021})}.
  \bibinfo{pages}{119–130}.
\newblock
\urldef\tempurl%
\url{https://doi.org/10.1145/3446804.3446843}
\showDOI{\tempurl}


\bibitem[\protect\citeauthoryear{Roemer, Buchanan, Shacham, and Savage}{Roemer
  et~al\mbox{.}}{2012}]%
        {roemer2012return}
\bibfield{author}{\bibinfo{person}{Ryan Roemer}, \bibinfo{person}{Erik
  Buchanan}, \bibinfo{person}{Hovav Shacham}, {and} \bibinfo{person}{Stefan
  Savage}.} \bibinfo{year}{2012}\natexlab{}.
\newblock \showarticletitle{Return-oriented programming: Systems, languages,
  and applications}.
\newblock \bibinfo{journal}{\emph{ACM Transactions on Information and System
  Security (TISSEC)}} \bibinfo{volume}{15}, \bibinfo{number}{1}
  (\bibinfo{year}{2012}), \bibinfo{pages}{2}.
\newblock


\bibitem[\protect\citeauthoryear{Scheifler}{Scheifler}{1977}]%
        {Scheifler1977AAIS}
\bibfield{author}{\bibinfo{person}{Robert~W. Scheifler}.}
  \bibinfo{year}{1977}\natexlab{}.
\newblock \showarticletitle{An Analysis of Inline Substitution for a Structured
  Programming Language}.
\newblock \bibinfo{journal}{\emph{Commun. ACM}} \bibinfo{volume}{20},
  \bibinfo{number}{9} (\bibinfo{date}{Sep.} \bibinfo{year}{1977}),
  \bibinfo{pages}{647–654}.
\newblock
\showISSN{0001-0782}
\urldef\tempurl%
\url{https://doi.org/10.1145/359810.359830}
\showURL{%
\tempurl}


\bibitem[\protect\citeauthoryear{Sehr, Muth, Biffle, Khimenko, Pasko, Schimpf,
  Yee, and Chen}{Sehr et~al\mbox{.}}{2010}]%
        {Sehr2010ASF}
\bibfield{author}{\bibinfo{person}{David Sehr}, \bibinfo{person}{Robert Muth},
  \bibinfo{person}{Cliff Biffle}, \bibinfo{person}{Victor Khimenko},
  \bibinfo{person}{Egor Pasko}, \bibinfo{person}{Karl Schimpf},
  \bibinfo{person}{Bennet Yee}, {and} \bibinfo{person}{Brad Chen}.}
  \bibinfo{year}{2010}\natexlab{}.
\newblock \showarticletitle{Adapting Software Fault Isolation to Contemporary
  CPU Architectures}. In \bibinfo{booktitle}{\emph{19th USENIX Conference on
  Security (USENIX)}}. \bibinfo{address}{Washington, DC}.
\newblock


\bibitem[\protect\citeauthoryear{{Shoshitaishvili}, {Wang}, {Salls},
  {Stephens}, {Polino}, {Dutcher}, {Grosen}, {Feng}, {Hauser}, {Kruegel}, and
  {Vigna}}{{Shoshitaishvili} et~al\mbox{.}}{2016}]%
        {Angr2016}
\bibfield{author}{\bibinfo{person}{Y. {Shoshitaishvili}}, \bibinfo{person}{R.
  {Wang}}, \bibinfo{person}{C. {Salls}}, \bibinfo{person}{N. {Stephens}},
  \bibinfo{person}{M. {Polino}}, \bibinfo{person}{A. {Dutcher}},
  \bibinfo{person}{J. {Grosen}}, \bibinfo{person}{S. {Feng}},
  \bibinfo{person}{C. {Hauser}}, \bibinfo{person}{C. {Kruegel}}, {and}
  \bibinfo{person}{G. {Vigna}}.} \bibinfo{year}{2016}\natexlab{}.
\newblock \showarticletitle{SOK: (State of) The Art of War: Offensive
  Techniques in Binary Analysis}. In \bibinfo{booktitle}{\emph{2016 IEEE
  Symposium on Security and Privacy (SP)}}. \bibinfo{address}{San Jose, CA,
  USA}.
\newblock


\bibitem[\protect\citeauthoryear{v.~d. Veen, Göktas, Contag, Pawoloski, Chen,
  Rawat, Bos, Holz, Athanasopoulos, and Giuffrida}{v.~d. Veen
  et~al\mbox{.}}{2016}]%
        {vanderVeen2016ATC}
\bibfield{author}{\bibinfo{person}{V. v.~d. Veen}, \bibinfo{person}{E.
  Göktas}, \bibinfo{person}{M. Contag}, \bibinfo{person}{A. Pawoloski},
  \bibinfo{person}{X. Chen}, \bibinfo{person}{S. Rawat}, \bibinfo{person}{H.
  Bos}, \bibinfo{person}{T. Holz}, \bibinfo{person}{E. Athanasopoulos}, {and}
  \bibinfo{person}{C. Giuffrida}.} \bibinfo{year}{2016}\natexlab{}.
\newblock \showarticletitle{A Tough Call: Mitigating Advanced Code-Reuse
  Attacks at the Binary Level}. In \bibinfo{booktitle}{\emph{2016 IEEE
  Symposium on Security and Privacy (SP)}}. \bibinfo{address}{San Jose, CA,
  USA}.
\newblock


\bibitem[\protect\citeauthoryear{van~der Veen, Andriesse, G\"{o}kta\c{s}, Gras,
  Sambuc, Slowinska, Bos, and Giuffrida}{van~der Veen et~al\mbox{.}}{2015}]%
        {vanderVeen2015PCC}
\bibfield{author}{\bibinfo{person}{Victor van~der Veen},
  \bibinfo{person}{Dennis Andriesse}, \bibinfo{person}{Enes G\"{o}kta\c{s}},
  \bibinfo{person}{Ben Gras}, \bibinfo{person}{Lionel Sambuc},
  \bibinfo{person}{Asia Slowinska}, \bibinfo{person}{Herbert Bos}, {and}
  \bibinfo{person}{Cristiano Giuffrida}.} \bibinfo{year}{2015}\natexlab{}.
\newblock \showarticletitle{Practical Context-Sensitive CFI}. In
  \bibinfo{booktitle}{\emph{22nd ACM SIGSAC Conference on Computer and
  Communications Security (CCS)}}. \bibinfo{address}{Denver, Colorado, USA}.
\newblock


\bibitem[\protect\citeauthoryear{Wang, Shoshitaishvili, Bianchi, Machiry,
  Grosen, Grosen, Kruegel, and Vigna}{Wang et~al\mbox{.}}{2017}]%
        {wang2017ramblr}
\bibfield{author}{\bibinfo{person}{Ruoyu Wang}, \bibinfo{person}{Yan
  Shoshitaishvili}, \bibinfo{person}{Antonio Bianchi}, \bibinfo{person}{Aravind
  Machiry}, \bibinfo{person}{John Grosen}, \bibinfo{person}{Paul Grosen},
  \bibinfo{person}{Christopher Kruegel}, {and} \bibinfo{person}{Giovanni
  Vigna}.} \bibinfo{year}{2017}\natexlab{}.
\newblock \showarticletitle{Ramblr: Making reassembly great again}. In
  \bibinfo{booktitle}{\emph{24th Annual Symposium on Network and Distributed
  System Security (NDSS)}}. \bibinfo{address}{San Diego, CA, USA}.
\newblock


\bibitem[\protect\citeauthoryear{Welton and Miller}{Welton and Miller}{2019}]%
        {Welton2019Diogenes}
\bibfield{author}{\bibinfo{person}{Benjamin Welton} {and}
  \bibinfo{person}{Barton~P. Miller}.} \bibinfo{year}{2019}\natexlab{}.
\newblock \showarticletitle{Diogenes: Looking for an Honest CPU/GPU Performance
  Measurement Tool}. In \bibinfo{booktitle}{\emph{Proceedings of the
  International Conference for High Performance Computing, Networking, Storage
  and Analysis}} (Denver, Colorado) \emph{(\bibinfo{series}{SC'19})}.
\newblock


\bibitem[\protect\citeauthoryear{Welton and Miller}{Welton and Miller}{2020}]%
        {Welton2020IRP}
\bibfield{author}{\bibinfo{person}{Benjamin Welton} {and}
  \bibinfo{person}{Barton~P. Miller}.} \bibinfo{year}{2020}\natexlab{}.
\newblock \showarticletitle{Identifying and (Automatically) Remedying
  Performance Problems in CPU/GPU Applications}. In
  \bibinfo{booktitle}{\emph{34th ACM International Conference on Supercomputing
  (ICS)}}. \bibinfo{address}{Barcelona, Spain}, Article
  \bibinfo{articleno}{27}, \bibinfo{numpages}{13}~pages.
\newblock


\bibitem[\protect\citeauthoryear{Williams-King, Kobayashi, Williams-King,
  Patterson, Spano, Wu, Yang, and Kemerlis}{Williams-King
  et~al\mbox{.}}{2020}]%
        {WilliamsKing2020Egalito}
\bibfield{author}{\bibinfo{person}{David Williams-King},
  \bibinfo{person}{Hidenori Kobayashi}, \bibinfo{person}{Kent Williams-King},
  \bibinfo{person}{Graham Patterson}, \bibinfo{person}{Frank Spano},
  \bibinfo{person}{Yu~Jian Wu}, \bibinfo{person}{Junfeng Yang}, {and}
  \bibinfo{person}{Vasileios~P. Kemerlis}.} \bibinfo{year}{2020}\natexlab{}.
\newblock \showarticletitle{Egalito: Layout-Agnostic Binary Recompilation}. In
  \bibinfo{booktitle}{\emph{Twenty-Fifth International Conference on
  Architectural Support for Programming Languages and Operating Systems
  (ASPLOS)}}. \bibinfo{address}{Lausanne, Switzerland}.
\newblock


\bibitem[\protect\citeauthoryear{Williams-King and Yang}{Williams-King and
  Yang}{2019}]%
        {WilliamsKing2019CodeMason}
\bibfield{author}{\bibinfo{person}{David Williams-King} {and}
  \bibinfo{person}{Junfeng Yang}.} \bibinfo{year}{2019}\natexlab{}.
\newblock \showarticletitle{CodeMason: Binary-Level Profile-Guided
  Optimization}. In \bibinfo{booktitle}{\emph{3rd ACM Workshop on Forming an
  Ecosystem Around Software Transformation}} (London, United Kingdom)
  \emph{(\bibinfo{series}{FEAST'19})}.
\newblock


\bibitem[\protect\citeauthoryear{Wojtczuk}{Wojtczuk}{2001}]%
        {wojtczuk2001advanced}
\bibfield{author}{\bibinfo{person}{Rafal Wojtczuk}.}
  \bibinfo{year}{2001}\natexlab{}.
\newblock \showarticletitle{The advanced return-into-lib (c) exploits: PaX case
  study}.
\newblock \bibinfo{journal}{\emph{Phrack Magazine, Volume 0x0b, Issue 0x3a,
  Phile\# 0x04 of 0x0e}} (\bibinfo{year}{2001}).
\newblock


\bibitem[\protect\citeauthoryear{Xu, Wang, Song, John, and Liu}{Xu
  et~al\mbox{.}}{2019b}]%
        {Xu2019CWT}
\bibfield{author}{\bibinfo{person}{Hao Xu}, \bibinfo{person}{Qingsen Wang},
  \bibinfo{person}{Shuang Song}, \bibinfo{person}{Lizy~Kurian John}, {and}
  \bibinfo{person}{Xu Liu}.} \bibinfo{year}{2019}\natexlab{b}.
\newblock \showarticletitle{Can We Trust Profiling Results? Understanding and
  Fixing the Inaccuracy in Modern Profilers}. In
  \bibinfo{booktitle}{\emph{Proceedings of the ACM International Conference on
  Supercomputing}} (Phoenix, Arizona, USA) \emph{(\bibinfo{series}{ICS '19})}.
  \bibinfo{pages}{284–295}.
\newblock
\urldef\tempurl%
\url{https://doi.org/10.1145/3330345.3330371}
\showDOI{\tempurl}


\bibitem[\protect\citeauthoryear{Xu, Ghaffarinia, Wang, Hamlen, and Lin}{Xu
  et~al\mbox{.}}{2019a}]%
        {Xu2019CONFIRM}
\bibfield{author}{\bibinfo{person}{Xiaoyang Xu}, \bibinfo{person}{Masoud
  Ghaffarinia}, \bibinfo{person}{Wenhao Wang}, \bibinfo{person}{Kevin~W.
  Hamlen}, {and} \bibinfo{person}{Zhiqiang Lin}.}
  \bibinfo{year}{2019}\natexlab{a}.
\newblock \showarticletitle{{CONFIRM}: Evaluating Compatibility and Relevance
  of Control-flow Integrity Protections for Modern Software}. In
  \bibinfo{booktitle}{\emph{28th {USENIX} Security Symposium ({USENIX}
  Security)}} (Santa Clara, CA).
\newblock


\bibitem[\protect\citeauthoryear{Yi, Dong, Dong, and Chen}{Yi
  et~al\mbox{.}}{2020}]%
        {Yi2020PEBS}
\bibfield{author}{\bibinfo{person}{Jifei Yi}, \bibinfo{person}{Benchao Dong},
  \bibinfo{person}{Mingkai Dong}, {and} \bibinfo{person}{Haibo Chen}.}
  \bibinfo{year}{2020}\natexlab{}.
\newblock \showarticletitle{On the Precision of Precise Event Based Sampling}.
  In \bibinfo{booktitle}{\emph{Proceedings of the 11th ACM SIGOPS Asia-Pacific
  Workshop on Systems}} (Tsukuba, Japan) \emph{(\bibinfo{series}{APSys '20})}.
  \bibinfo{pages}{98–105}.
\newblock
\showISBNx{9781450380690}
\urldef\tempurl%
\url{https://doi.org/10.1145/3409963.3410490}
\showDOI{\tempurl}


\bibitem[\protect\citeauthoryear{Zhang and Sekar}{Zhang and Sekar}{2013}]%
        {Zhang2013CFI}
\bibfield{author}{\bibinfo{person}{Mingwei Zhang} {and} \bibinfo{person}{R.
  Sekar}.} \bibinfo{year}{2013}\natexlab{}.
\newblock \showarticletitle{Control Flow Integrity for COTS Binaries}. In
  \bibinfo{booktitle}{\emph{Presented as part of the 22nd USENIX Security
  Symposium (USENIX Security 13)}} (Washington, D.C.).
\newblock


\bibitem[\protect\citeauthoryear{Zhu, Feng, Meng, Wen, Camtepe, Xiang, and
  Ren}{Zhu et~al\mbox{.}}{2020}]%
        {Zhu2020CSIFuzz}
\bibfield{author}{\bibinfo{person}{Xiaogang Zhu}, \bibinfo{person}{Xiaotao
  Feng}, \bibinfo{person}{Xiaozhu Meng}, \bibinfo{person}{Sheng Wen},
  \bibinfo{person}{Seyit Camtepe}, \bibinfo{person}{Yang Xiang}, {and}
  \bibinfo{person}{Kui Ren}.} \bibinfo{year}{2020}\natexlab{}.
\newblock \showarticletitle{CSI-Fuzz: Full-speed Edge Tracing Using Coverage
  Sensitive Instrumentation}.
\newblock \bibinfo{journal}{\emph{IEEE Transactions on Dependable and Secure
  Computing}} (\bibinfo{year}{2020}), \bibinfo{pages}{1--1}.
\newblock
\urldef\tempurl%
\url{https://doi.org/10.1109/TDSC.2020.3008826}
\showDOI{\tempurl}


\end{thebibliography}
